\documentclass[useAMS,usenatbib]{mn2e}

\usepackage{graphicx}
\usepackage{grffile}\usepackage{epsf}
\usepackage{amsmath}
\usepackage{float}
\usepackage{fixltx2e}

\usepackage{setspace}
\usepackage{mathrsfs}

\newcommand{\kmsmpc}{\kms\;{\rm Mpc}^{-1}}

\newcommand{\hkpc}{h^{-1}{\rm kpc}}

\newcommand{\kms}{{\rm km}\,{\rm s}^{-1}}
\newcommand{\cms}{{\rm cm}^{-2}}
\newcommand{\cmc}{{\rm cm}^{-3}}

\newcommand{\msolar}{{\rm M}_{\odot}}
\newcommand{\msolaryr}{{\rm M}_{\odot} {\rm yr}^{-1}}

\newcommand{\gad}{{\sc Gadget-3}}

\newcommand{\CIV}{\hbox{C\,{\sc iv}}}

\newcommand{\SiII}{{\hbox{Si\,{\sc ii}}}}
\newcommand{\SiIII}{{\hbox{Si\,{\sc iii}}}}
\newcommand{\SiIV}{\hbox{Si\,{\sc iv}}}

\newcommand{\NV}{\hbox{N\,{\sc v}}}

\newcommand{\OIII}{\hbox{O\,{\sc iii}}}

\newcommand{\OVI}{\hbox{O\,{\sc vi}}}
\newcommand{\OVII}{\hbox{O\,{\sc vii}}}

\newcommand{\OIX}{\hbox{O\,{\sc ix}}}
\newcommand{\HI}{{\hbox{H\,{\sc i}}}}

\newcommand{\sdotMBH}{{{\rm s}\dot{M}_{\rm BH}}}
\newcommand{\tAGN}{t_{\rm AGN}}
\newcommand{\ergs}{{\rm erg}\,{\rm s}^{-1}}

\newcommand{\nh}{{n_{\rm H}}}
\newcommand{\apjl}{{ApJ}}

\newcommand{\apj}{{ApJ}}
\newcommand{\apjs}{{ApJS}}

\newcommand{\mnras}{{MNRAS}}

\begin{document}
\title[Flickering AGN can explain strong O VI]{Flickering AGN can explain the strong circumgalactic O VI observed by COS-Halos}

\author[B. D. Oppenheimer et al.]{
\parbox[t]{\textwidth}{\vspace{-1cm}
  Benjamin D. Oppenheimer$^{1}$\thanks{benjamin.oppenheimer@colorado.edu}, Marijke Segers$^{2}$, Joop~Schaye$^{2}$, Alexander J. Richings$^{3}$, Robert A. Crain$^{4}$}\\\\
$^1$CASA, Department of Astrophysical and Planetary Sciences, University of Colorado, 389 UCB, Boulder, CO 80309, USA\\
$^2$Leiden Observatory, Leiden University, P.O. Box 9513, 2300 RA, Leiden, The Netherlands\\
$^3$Department of Physics and Astronomy and CIERA, Northwestern University, 2145 Sheridan Road, Evanston, IL 60208, USA\\
$^4$Astrophysics Research Institute, Liverpool John Moores University, 146 Brownlow Hill, Liverpool, L3 5RF, UK\\
}
\maketitle

\pubyear{2017}

\maketitle

\label{firstpage}

\begin{abstract}

Proximity zone fossils (PZFs) are ionization signatures around
recently active galactic nuclei (AGN) where metal species in the
circumgalactic medium remain over-ionized after the AGN has shut-off
due to their long recombination timescales.  We explore cosmological
zoom hydrodynamic simulations using the EAGLE model paired with a
non-equilibrium ionization and cooling module including time-variable
AGN radiation to model PZFs around star-forming, disk galaxies in the
$z\sim 0.2$ Universe.  Previous simulations typically under-estimated
the $\OVI$ content of galactic haloes, but we show that plausible PZF
models increase $\OVI$ column densities by $2-3\times$ to achieve the
levels observed around COS-Halos star-forming galaxies out to 150 kpc.
Models with AGN bolometric luminosities $\ga 10^{43.6} \ergs$, duty
cycle fractions $\la 10\%$, and AGN lifetimes $\la 10^6$ yr are the
most promising, because their super-massive black holes grow at the
cosmologically expected rate and they mostly appear as inactive AGN,
consistent with COS-Halos.  The central requirement is that the
typical star-forming galaxy hosted an active AGN within a timescale
comparable to the recombination time of a high metal ion, which for
circumgalactic $\OVI$ is $\approx 10^7$ years.  $\HI$, by contrast,
returns to equilibrium much more rapidly due to its low neutral
fraction and does not show a significant PZF effect.  $\OVI$
absorption features originating from PZFs appear narrow, indicating
photo-ionization, and are often well-aligned with lower metal ion
species.  PZFs are highly likely to affect the physical interpretation
of circumgalactic high ionization metal lines if, as expected, normal
galaxies host flickering AGN.


\end{abstract}


\section{Introduction}  

The circumgalactic medium (CGM) contains a significant reservoir of
gaseous baryons extending to the virial radius and beyond.
Observations by the Cosmic Origins Spectrograph (COS) show the gas is
enriched with heavy elements out to galactocentric radii of at least
150 kpc, as indicated by metal ion absorption features including
$\OVI$.  By targeting the CGM of star-forming, redshift $z\sim 0.2$
galaxies, the COS-Halos survey found very strong $\OVI$, with an
average column density $N_{\OVI}=10^{14.6} \cms$ \citep{tum11}.

The estimated circumgalactic reservoir of $\OVI$, in excess of
$2\times 10^6 \msolar$ \citep{pee14}, indicates significant enrichment
by galactic superwinds.  Cosmological hydrodynamic simulations have
been unable to reproduce the typical $\OVI$ columns observed around
star-forming galaxies.  The latest simulations confronting COS-Halos,
whether using smoothed-particle hydrodynamics \citep[SPH;][]{for16,
  gut17}, adaptive mesh refinement \citep{hum13}, or moving mesh
\citep{sur17}, all seem to fall short by a factor of $\approx
3$. These simulations generate $\OVI$ column densities of $\approx
10^{14.0} \cms$ around star-forming, $L^*$ galaxies with stellar
masses $M_*= 10^{10.0}-10^{10.5} \msolar$.

\citet[][hereafter Opp16]{opp16a} also ran a set of zoom simulations
using the EAGLE \citep[Evolution and Assembly of GaLaxies and their
  Environments;][]{sch15} model, reproducing the observed bimodal
correlation of $\OVI$ column density with galactic specific star
formation rate \citep[sSFR;][]{tum11}.  While these were the first
cosmological hydrodynamic simulations of galaxies to integrate
non-equilibrium (NEQ) ionization and cooling following 136 ions across
11 elements \citep{opp13a}, they too find $\OVI$ was too weak around
$L^*$ galaxies.  The NEQ ion-by-ion cooling and abundances under a
slowly evolving, spatially uniform ionizing extra-galactic background
(EGB) does not significantly affect $\OVI$ column densities.  Hence,
circumgalactic $\OVI$ in this case is well approximated as being in
ionization equilibrium and is found to be too weak.

However, Opp16 did not consider the case of a fluctuating ionization
field, such as that arising from a time-variable active galactic
nucleus (AGN).  \citet[][hereafter OS13]{opp13b} introduced the
concept of AGN proximity zone fossils (or PZFs), where an AGN ionizes
circumgalactic gas and subsequently turns off, leaving metals in the
CGM over-ionized for a timescale set by their recombination rate.
\citet[][herafter Seg17]{seg17} were the first paper to simulate PZFs
in a cosmological hydro simulation.  Using individual haloes selected
from the EAGLE simulation, Seg17 performed a parameter exploration of
AGN strength and lifetime, duty cycle, halo mass, and redshift,
finding that the PZF effect could maintain the column densities of
high ions (including $\OVI$) significantly above the equilibrium
values for much longer than the AGN lifetime.  {\it A galaxy's CGM
  would appear ionized by an AGN despite the central galaxy showing no
  signatures of AGN activity.}

The key requirement for the PZF effect to be significant is that the
recombination timescale ($t_{\rm rec}$) to the observed ionization
state should be similar to, or longer than, the time between AGN
episodes.  For a typical CGM density, $\nh=10^{-4} \cmc$, the
recombination time to $\OVI$ is $t_{\rm rec} \sim 10^7$ yr.  This may
seem short compared to the $\HI$ recombination time of $10^9$ yr at
the same density, but the critical difference is that circumgalactic
hydrogen is highly ionized with a neutral fraction $\approx 10^{-4}$,
which reduces the timescale to re-equilibrate to $\approx 10^{-4}
\times t_{{\rm rec},\HI} \sim 10^5$ yrs.  By contrast, metal ions have
fractions for individual ions of order unity, so the full recombination
timescale applies.  Therefore, the traditional method of identifying a
proximity zone using ionized $\HI$ around an active quasar
\citep[e.g.][]{sco00} does not apply for PZFs.  OS13 argued, and Seg17
demonstrated, that PZFs should be identified as having $\HI$ levels
consistent with being in equilibrium with the EGB, while metal ions
still show signatures of being ionized by the AGN.

Here, we extend the work of Seg17 to argue that PZFs could be common
around COS-Halos $z\sim 0.2$ star-forming galaxies.  The COS-Halos
sample includes no active AGN \citep{wer12} while circumgalactic $\HI$
is copious \citep{tho12, tum13}.  Our argument for $\OVI$ enhanced by
PZFs hinges on the assumption that many of these galaxies were AGN
within the recent past, defined roughly as $t_{{\rm rec}, \OVI}$.  AGN
lifetimes are typically inferred to be $t_{\rm AGN}\sim 10^6-10^8$ yr
\citep[e.g.][]{hai01,mar01,jak03,schi04,hopk06,gon08}, but these
measurements constrain duty cycle, and individual AGN-on episodes
could be much shorter.  We additionally explore $t_{\rm AGN}\sim 10^5$
yr here, which is motivated by arguments including those of
\citet{gab13} and \citet{schaw15} that super-massive black holes
(SMBH) flicker with shorter, luminous AGN-on phases.

OS13 showed that the most important parameter for PZF enhancement is
the average interval between AGN episodes, which is a combination of
$t_{\rm AGN}$ and duty cycle fraction ($f_{\rm duty}$), or
approximately $t_{\rm AGN}/f_{\rm duty}$.  Seg17 demonstrated that
short, flickering AGN-on phases create stronger PZFs effects than a
single, longer AGN-on phase, because $t_{rec}$ remains the same as
long as the AGN can ionize the CGM within a single burst.
Additionally, subsequent AGN on-phases can accumulate to raise the
ionization state of the CGM, further enhancing the PZF effect (OS13,
Seg17).  Hence the PZF effect can non-linearly enhance metal ions in
the CGM if an AGN flickers.

AGN light echoes, where an AGN recently shut off or became obscured
along the line of sight to the object but leave the surrounding CGM at
an enhanced ionization level, may be observations of on-going PZF
effects.  \citet{lin09} found enhanced [$\OIII$] out to 27 kpc from IC
2497, discovered in Galaxy Zoo as ``Hanny's Voorwerp'' and suggesting
a change in AGN activity within the last $10^5$ yr.  The enhanced
oxygen ionization may be a direct example of the PZF effect, albeit in
higher density gas that can be detected in emission.  \citet{bla13}
argued that a 'Seyfert flare' in our own Galaxy $1-3$ Myr ago, which
also formed the {\it Fermi} bubble observed from Sgr A$^*$, could be
responsible for the strong H$\alpha$ observed in the Magellanic
Stream.  \citet{schir16} discovered 14 Lyman-$\alpha$ blobs at
$z\approx 0.3$ that appear to be associated with flickering AGN that
have faded in the X-ray by a factor of $10^{3-4}$ in the last
$10^{4-5}$ yr.  These three examples may be PZF effects that are
detected in emission, but our exploration focuses on quasar absorption
lines probing more diffuse, more highly ionized CGM gas.

The paper is organized as follows.  We describe our zoom simulations
and the implementation of AGN fluctuations into the non-equilibrium
ionization module included in the \gad-based EAGLE code in
\S\ref{sec:sims}.  We next consider the latest observational and
theoretical results regarding how SMBHs grow in low-redshift
star-forming galaxies in \S\ref{sec:design}.  Our main results are
presented in \S\ref{sec:results}, and we discuss their implications in
\S\ref{sec:discuss}.  We summarize in \S\ref{sec:summary}.

\section{Simulations} \label{sec:sims}  

We use the EAGLE simulation code described by \citet[][hereafter
Sch15]{sch15} and \citet{cra15}, which is an extensively modified
version of the N-body+SPH code \gad~last described by \citet{spr05}.
EAGLE includes calibrated prescriptions for star formation, stellar
evolution and chemical enrichment, and superwind feedback associated
with star formation and the SMBH growth.  Because EAGLE successfully
reproduces an array of stellar and cold ISM properties of galaxies
across a Hubble time
\citep[e.g. Sch15;][]{fur15,lag15,tra15,bah16,seg16a,cra17}, while
following the hydrodynamics, it represents an ideal testbed for the
study of the physical state of the gaseous intergalactic medium (IGM)
and the CGM.  Absorption line statistics probing the IGM are examined
by \citet{rah15a} for $\HI$ and by \citet{rah16} for metal ions,
showing broad agreement with observations, although notably an
under-estimate of $\OVI$ absorber frequency for
$N_{\OVI} \ga 10^{14.5} \cms$ at $z\approx 0-0.5$.  \citet{ros16}
explored AGN and SMBH statistics in EAGLE finding excellent agreement
with observations between $z=0-1$, and \citet{mca17} demonstrated
EAGLE reproduces the observed relationships between galaxy star
formation rate (SFR) and SMBH accretion rate.  The underestimates of
strong $\OVI$ absorbers in \citet{rah16} and Opp16 provide motivation
for this work.

\subsection{Non-equilibrium network} \label{sec:NEQ}

Opp16 ran a set of high-resolution EAGLE zooms focusing on the CGM of
$z\sim 0.2$ galaxies similar to those observed by COS-Halos.  These
simulations include the NEQ solver introduced by \citet{opp13a}, and
integrated into the EAGLE code as described in Opp16.  The NEQ module
explicitly solves the reaction network of 136 ionization states for 11
elements that contribute most significantly to radiative cooling.  The
module replaces the equilibrium cooling module of \citet{wie09a},
which enables a self-consistent transition from equilibrium to NEQ
runs at a specified redshift.  The reaction network includes radiative
and di-electric recombination, collisional ionization,
photo-ionization, Auger ionization, and charge transfer, as well as
ion-by-ion cooling \citep{gna12,opp13a}.  The NEQ module is activated
only at late times ($z=0.5$ or $z=0.28$ in the Opp16 zooms), because
of its significant computational expense.  Opp16 follow the NEQ
network in all non-star-forming gas (i.e. the IGM/CGM), but do not
follow the ISM non-equilibrium behaviour and chemistry described by
\citet{ric14a}.  We also do not follow the effects of AGN radiation on
the ISM.

In Opp16, the NEQ effects were found to be minimal for $\OVI$ CGM
column densities under a slowly evolving EGB field, modeled using the
\citet{haa01} quasar+galaxy background.  Additionally, the dynamics of
the gas and the appearance of the galaxies did not differ
significantly between NEQ and equilibrium runs.

\subsection{Fluctuating AGN implementation} \label{sec:AGNmethod} 

Here we add to the NEQ module the capability of adding a spatially and
temporally variable ionizing field, corresponding to localized
ionization by an AGN.  We assume that the \citet{haa01} EGB is always
present, and add the ionizing spectrum derived by \citet{saz04} for
unobscured AGN.  This is the same spectrum used by Seg17, and shown in
their Figure 3.  We have calculated the photo-ionization rates,
$\Gamma_{x_i, \rm AGN}$, for ionization state $i$ of atomic species
$x$ for the Sazonov spectrum using

\begin{equation} \label{eqn:gamma}
\Gamma_{x_i, \rm AGN} = \int^{\infty}_{\nu_{0,x_i}} \frac{f_{\nu}}{h \nu} \sigma_{x_i}(\nu) d\nu,
\end{equation}

\noindent where $\nu$ is frequency, $\nu_{0,x_i}$ is the ionisation
frequency, $f_{\nu}$ is the flux from the AGN, $\sigma_{x_i}(\nu)$ is
photo-ionisation cross-section, and $h$ is the Planck constant.  Our
NEQ module uses the Sundials {\tt
  CVODE}\footnote{https://computation.llnl.gov/casc/sundials/main.html}
solver to integrate the ionization balance over a hydrodynamic
timestep, according to

\begin{equation}
\begin{split}
\frac{dn_{x_i}}{dt}& = n_{x_{i+1}} \alpha_{x_{i+1}} n_e + n_{x_{i-1}}(\beta_{x_{i-1}} n_e + 
\Gamma_{x_{i-1,}\rm EGB} \\
& + \Gamma_{x_{i-1},\rm AGN}) - n_{x_i} ((\alpha_{x_i}+\beta_{x_i}) n_e + \Gamma_{x_i,\rm EGB} + \Gamma_{x_i,\rm AGN}),
\end{split}
\end{equation}

\noindent where $n$ is the particle number density (cm$^{-3}$) for a
given $x_{i}$ ionization state, $n_e$ is the free electron density
(cm$^{-3}$), $\alpha_{x_i}$ is the total recombination rate
coefficient (radiative plus di-electric, cm$^3$ s$^{-1}$), and
$\beta_{x_i}$ is the collisional ionisation rate coefficient (cm$^3$
s$^{-1}$).

We modify the EAGLE \gad~code to calculate $\Gamma_{x_i, \rm AGN}$ for
each gas particle based on its distance from the central SMBH.  We do
not attempt to link AGN activity self-consistently to the simulated
accretion rate of the central SMBH.  Instead, we explore a variety of
AGN time histories based on several parameters including luminosity
($L_{\rm bol}$), AGN lifetime ($t_{\rm AGN}$, i.e. the length of time
the AGN is on), and duty cycle fraction ($f_{\rm duty}$).  We set
$t_{\rm AGN}$ by the timestep length over which all SPH particles are
updated, which is adjusted by setting the \gad~parameter for the
maximum hydro timestep.  A random number generator determines if the
AGN is on according to fraction $f_{\rm duty}$ at each such correlated
timestep.  This stochastic AGN flickering differs from the constant
interval between AGN episodes explored by Seg17.

A simulation with $t_{\rm AGN}=10^5$ yr takes a much longer run time
than one adopting $t_{\rm AGN}=10^7$ yr, because every particle is
updated on this short correlated timestep.  \gad~uses differential
timestepping and a normal run updates only the densest particles on a
$\la 10^5$ yr timestep at our zoom resolution.  The hydro performance
is essentially the same by resetting the maximum hydro timestep to a
smaller value, but more hydro timesteps are taken to track the
radiation field for shorter $t_{\rm AGN}$.  We explore timesteps of
$10^{5.0}$, $10^{6.2}$, and $10^{7.1}$ yr, which are slightly
out-of-sync with $1.0$ dex steps owing to \gad's time-stepping.

We do not perform radiative transfer calculations of the propagation
of AGN photons.  We simply turn on an AGN at a given timestep, and do
not consider light travel time effects, which we argue in
\S\ref{sec:discuss} is reasonable for our exploration.  While our main
models emit AGN radiation isotropically, we also explore biconical
opening angles for the AGN radiation.  Photo-heating from the AGN
radiation is included in these runs, although it is dynamically and
observationally unimportant as shown by OS13 and Seg17.  These are the
first dynamic simulation runs we know of with NEQ ionization and
cooling including fluctuating AGN.  \citet{vog13} integrated radiative
AGN feedback into AREPO simulations, but their assumption of
ionization equilibrium precludes the PZF effect.

Finally, we note that our method of using dynamic EAGLE zoom runs
recovers the same results as the Seg17 method of using a halo from a
static snapshot.  For this test, we applied both methods to a $z=0.1$
zoom of a $M_* = 2\times 10^{10} \msolar$ galaxy, and verified that 1)
$\OVI$, $\CIV$, and $\HI$ as functions of impact parameter and time
agreed, and 2) the ionization histories for individual gas particles
agreed.

\subsection{Zoom simulations} 

The zoom simulation method is presented in \S2.3 of Opp16, and we
briefly describe some details here.  We use the zoom initial condition
generation methods described by \citet{jen10, jen13a}.  \citet{pla14}
cosmological parameters as adopted by EAGLE are $\Omega_{\rm
  m}=0.307$, $\Omega_{\Lambda}=0.693$, $\Omega_b=0.04825$, $H_0=
67.77$ $\kmsmpc$, $\sigma_8=0.8288$, and $n_{\rm s}=0.9611$.  We
choose one target halo for resimulation from the EAGLE Recal-L025N0752
simulation, and particles are identified at $z=0$ in a spherical
region with radius $3 R_{200}$ (where $R_{200}$ encloses an
overdensity of $200\times$ the critical overdensity).

We select the Gal001 halo listed in Table 1 of Opp16 for our
exploration of the AGN PZF effect.  In contrast to Opp16, who used a
set of 20 zoom simulations to simulate the star-forming and passive
COS-Halos sample, we limit our exploration to a single zoom applied to
the star-forming sample.  This zoom shows typical $\OVI$ columns of
the Opp16 star-forming blue galaxy sample (see \S\ref{sec:results}),
meaning we can use it as the baseline comparison for the PZF effect.
The primary zoom resolution is {\it M5.3}, where the resolution
nomenclature refers to SPH particle mass resolution according to {\it
  M}[log($m_{\rm SPH}/\msolar$)].  At $z=0.2$, the halo mass is
$M_{200}=10^{12.07} \msolar$, the central galaxy stellar mass is
$M_{*}=10^{10.27} \msolar$, and the SFR is $1.4 \msolaryr$ in the run
without radiative AGN activity.  The softening is 350 proper pc at
$z<2.8$.  This zoom simulates a typical star-forming, $L^*$ galaxy
using the EAGLE-``Recal'' prescription described by Sch15.  This
object is an intermediate mass $L^*$ galaxy from Opp16, and its
properties, as well as the EAGLE-Recal prescription feedback
parameters, are listed in Table \ref{tab:zooms}.  The run without AGN
radiative activity from Opp16 is termed the No-AGN-Rad run.

\begin{table*}
\caption{Zoom simulation runs with flickering AGN}
\begin{tabular}{ccrrrrrrr}
\hline
Name & 
Resolution$^a$ &
$f_{\rm th,max}$ &
$f_{\rm th,min}$ &
$C_{\rm visc}/2\pi$ &
log $M_{200}^b$ ($\msolar$) &
log $M_{*}^b$ ($\msolar$) &
SFR$^b$ ($\msolaryr$) &
$M_{\rm BH}^b$ ($\msolar$) 
\\
\hline
\multicolumn {8}{c}{}\\
Gal001  & {\it M5.3} & 3.0 & 0.3 & $10^3$ & 12.07 & 10.27 & 1.426 & $10^{7.06}$\\
Gal001  & {\it M4.4} & 2.0 & 0.3 & $10^4$ & 12.09 & 10.28 & 1.791 & $10^{6.85}$\\
\hline
\end{tabular}
\\
\parbox{25cm}{
$^a$ M[log$_{10}$($m_{\rm SPH}/\msolar$)], $^b$ At $z=0.205$
}
\label{tab:zooms}
\end{table*}

We also simulate the Gal001 halo at $8\times$ higher mass resolution,
{\it M4.4}, to test resolution convergence in Appendix \ref{sec:res}.
This is run in equilibrium from $z=127$ with a modified feedback
prescription to compensate for the fact that stellar masses at {\it
  M4.4} resolution were 0.18 dex lower using the EAGLE-Recal
prescription (Opp16).  The EAGLE strategy is to recalibrate the
subgrid models at different resolutions as described by Sch15, which
was not done in Opp16 for {\it M4.4} zooms.  Here, we take the
EAGLE-Recal prescription and reduce the stellar feedback energy by
lowering the parameter $f_{\rm th,max}$ of Sch15 from $3.0$ to $2.0$,
and reducing the viscosity of gas accreting onto the SMBH by raising
the $C_{\rm visc}$ parameter that scales inversely with subgrid
viscosity \citep{ros15} from $2\pi\times 10^3$ to $2\pi\times 10^4$.
The result is a galaxy with a similar stellar mass as the {\it M5.3}
run, and a similar albeit slightly lower mass black hole at the center
(see Table \ref{tab:zooms}).

Finally, we run a set of {\it M5.3} models with the AGN turned on to
demonstrate the expectations of constant AGN radiation on the CGM.
This is not applicable to COS-Halos, but could be applicable to a
survey targeting sight lines around active AGN, including the COS-AGN
survey (Berg et al., in prep).  We present the $\OVI$ results in the
Appendix \ref{sec:AGNon}.

\subsection{Running in non-equilibrium}

Simulations are run with non-equilibrium ionization and cooling
from $z=0.271$, and AGN fluctuations begin between $z=0.230$
and $0.225$.  Because we compare to the COS-Halos observations centered
at $z\sim 0.2$, we focus our AGN fluctuation snapshots around this
redshift, usually running simulations to $z=0.15$.

We parameterize AGN luminosity using $L_{\rm bol}$, exploring the
interval log[$L_{\rm bol}$/(erg s$^{-1}$)]$= 43.1-45.1$, which
corresponds to Eddington ratios,

\begin{equation}
\lambda_{\rm Edd}\equiv \frac{L_{\rm bol}}{L_{\rm Edd}}
\end{equation}

\noindent ranging between $10^{-2}$ and $1$ for $M_{\rm BH}=10^{7.0}
\msolar$.  We assume $10^{7.0} \msolar$ when quoting Eddington ratios,
and we do not adjust bolometric luminosities based on the SMBH mass,
which differs between {\it M5.3} and {\it M4.4} resolutions and
increases slightly over our explored redshift interval.  These SMBH
masses are typical in the EAGLE cosmological volumes where the median
SMBH mass is $10^{7} \msolar$ for a $10^{12} \msolar$ halo at $z=0.0$
\citep{ros16}.

We output high-cadence ``snipshots,'' which are abbreviated EAGLE
snapshots with fewer fields and only a handful of NEQ ions.  Snipshots
are output every 4 Myr for a total of $234$ between $z=0.23$ and
$0.15$.  For most of the $t_{\rm AGN}=10^5$ yr runs, we only run to
just below $z=0.20$ owing to the greater expense of these runs.
However there is no statistical difference in $\OVI$ columns in a
No-AGN-Rad run between $z=0.23\rightarrow0.19$ and
$z=0.23\rightarrow0.15$. 


\subsection{Observational sample selection}

We take the subset of the COS-Halos blue, star-forming galaxies
defined as having sSFR greater than $10^{-11}$ yr$^{-1}$ and stellar
masses between $10^{9.8}-10^{10.5} \msolar$ using a \citet{cha03} IMF.
This leaves $20$ galaxies with a median sSFR$=10^{-10.0}$ yr$^{-1}$
using the catalogue of \citet{wer13}.  The observed median $\OVI$
column density is $10^{14.57} \cms$ for impact parameters $b=20-140$
kpc.  The additional six galaxies in the COS-Halos blue sample with
$M_*>10^{10.5} \msolar$ have slightly lower sSFR and lower $\OVI$
columns including two upper limits below $10^{14.0} \cms$ and a median
value of $10^{14.38} \cms$.  The redshift range of the observed sample
is $z=0.14-0.36$, with a median of $z=0.22$.

Our simulated {\it M5.3} central galaxy with $M_* = 10^{10.3} \msolar$
and sSFR$=10^{-10.1}$ yr$^{-1}$ is representative of the COS-Halos
blue subset.  We explore a limited redshift range, but one which
includes the median redshift of the observed sample.  Opp16 found no
significant dependence of $\OVI$ column densities on redshift.


\section{Experiment design rationale} \label{sec:design}

Our hypothesis that the PZF effect boosts the COS-Halos $\OVI$ column
densities relies on the majority of COS-Halos star-forming galaxies
hosting AGN luminous enough to have ionized the CGM in the recent
past, which is defined as the lifetime of the PZF set by the
recombination timescale to the given metal ion.  OS13 and Seg17 show
that $t_{\rm rec}$ for $\OVI$ is $5-10$ Myr at typical $z\sim 0.2$ CGM
densities, which means that most of the COS-Halos blue sample should
have hosted a luminous AGN within this timescale.  Another requirement
is that the AGN are infrequent enough that it is statistically
plausible that none of the COS-Halos blue galaxies appear as active
AGN, because none of them appear as such according to \citet{wer12}.
Finally, the AGN luminosities, lifetimes, and duty cycles need to be
``evolutionarily sustainable,'' such that the implied SMBH growth
rates are reasonable for a typical star-forming galaxy, and therefore
cosmologically expected.  The hypothesis would fail if we required
many of the $20$ observed galaxies to be undergoing a rare stage of
SMBH growth.  We now consider recent observations and theoretical
results regarding AGN and SMBH growth in low-redshift, low-luminosity
AGN.

\subsection{AGN luminosities and duty cycles in star-forming galaxies}

Our typical SMBH mass is $10^7 \msolar$ in a $M_* = 10^{10.3} \msolar$
galaxy, and is representative of EAGLE cosmological volumes
\citep{ros16}.  The observed \citet{har04} SMBH mass-galaxy mass
relation predicts about $\approx 2\times$ more massive SMBH in their
local sample, but it should be noted that this is a bulge-dominated
sample.  Recent work shows that disk-dominated galaxies host
low-luminosity AGN, which is critical for our hypothesis since
COS-Halos star-forming galaxies are disk galaxies.  \citet{sun15}
compiled an X-ray-selected, {\it Herschel} cross-matched sample of
disk-dominated galaxies with AGN calculating a duty cycle of $\approx
10 \%$ with an average $\lambda_{\rm Edd}\sim 0.1$.

Eddington ratios range between $0.01-1$ for typical broad-line quasar
activity with the $\lambda_{\rm Edd}\sim 0.01$ lower limit observed by
\citet{kol06} and \citet{tru09}.  Eddington ratios approaching unity
are rare in the low-redshift Universe \citep[e.g.][]{she08}.  Such
high Eddington ratios are more likely to be associated with {\it
  quasar} activity, where an SMBH gains most of its mass in a couple
episodes \citep[e.g.][]{hopk09}, but we are considering {\it
  Seyfert}-level activity where lower Eddington ratios appear to
dominate SMBH growth.  \citet{hopk06b} developed a stochastic
cold-accretion model to fit the Seyfert luminosity function, where
``quiescent'' accretion dominates SMBH growth in spiral galaxies with
the most likely accretion luminosities having $\lambda_{\rm Edd}\sim
0.01$ and extending up to $\sim 0.1$ for $M_{\rm BH}=10^{7} \msolar$.

Low-luminosity AGN may not always appear as optically-selected AGN
using a method like the \citet[][BPT]{bal81} diagnostic as used by
\citet{wer12}.  \citet{sat14} demonstrated that optical selection may
miss many AGN in bulge-less galaxies, and that infrared {\it
  WISE}-selected AGN may be prevalent in star-forming galaxies.
\citet{tru15} showed that star formation can out-shine AGN activity in
high-sSFR disk galaxies.  It may be that much of the growth of $10^7
\msolar$ black holes is shielded from traditional optical
identification, although it is not clear what this means for the
far-UV spectrum, which is needed to ionize $\OVI$ in PZFs.

X-rays have fewer biases than optical detections, and the survey of
\citet{hag10} finds 1.2\% AGN detection fraction in field galaxies with
$L^*$ luminosities, which is consistent with COS-Halos showing no
active AGN.  However, in the real Universe, AGN are not on-off light
bulbs, and the intrinsic Eddington ratio distribution of star-forming
galaxy SMBHs may be best modeled by a power law distribution with a
steep cutoff at $\lambda_{\rm Edd}\ga 0.1$ \citep{jon16, air17}.

\subsection{AGN lifetimes in star-forming galaxies} \label{sec:des_lifetime}

Quasar lifetimes are calculated to be $\approx 10^6-10^8$ yr
\citep[e.g.][]{hai01,mar01} based on quasar clustering and halo
occupation, but these measurements constrain total duty cycle fraction
and not the duration of individual quasar episodes.  Ionization of the
surrounding IGM, often probed via the transverse proximity effect of
paired quasars, can also be used to indirectly constrain quasar
lifetimes to be $\ga 10^7$ yr \citep{jak03,schi04,gon08,bor16}, but
consideration of quasar flickering would require re-interpretation of
these results.
  
Recent results support shorter AGN lifetimes, especially when
considering low-luminosity AGN in disks.  \citet{schaw15} constrains
AGN lifetimes to $\approx 10^5$ yr based on the timing argument that
there is a lag between the AGN central engine becoming X-ray active
and ionizing its host galaxy's ISM as indicated in optical lines.
Hence, the fraction of X-ray bright, optically normal galaxies can be
combined with the time lag to argue that AGN flicker in many ($100-
1000$) accretion bursts.  The simulations of \citet{gab13} find the
accretion of smaller clouds in low-$z$ gas-poor disks leads to shorter
AGN active phases of $10^5$ yr, in contrast to longer phases in
high-$z$, gas-rich disks.  Other theoretical work considering
accretion on pc and sub-pc scales \citep{nov11, kin15} also indicate
chaotic accretion episodes lasting $\approx 10^5$ yr.

\subsection{SMBH growth rates in star-forming galaxies}  \label{sec:des_mdot}

When considering stochastic AGN activity in a sample of star-forming
galaxies, one needs to consider the time-averaged rate of growth of
the black hole in relation to the growth of the galaxy.  The sSFR of
the blue COS-Halos sample implies a stellar mass doubling time of
$10^{10}$ yr, which is much longer than the \citet{sal64} timescale,
$t_{\rm Sal}\equiv M_{\rm BH}/\dot{M}_{\rm Edd}=4\times 10^7$ yr,
required for an Eddington-limited BH to double in mass assuming a
radiative efficiency, $\epsilon_{\rm rad}=10\%$, where $\dot M_{\rm
  Edd}= L_{\rm Edd}/(\epsilon_{\rm rad} c^2)$.

What combinations of duty cycle fraction and sub-Eddington accretion
rates are reasonable for SMBH growth in star-forming disk galaxies?
\citet{hic14} demonstrate that a model where long-term SMBH accretion
rates correlate with star formation rates can explain observed AGN
statistics.  The BH tracks the galaxy growth on a long timescale,
which means the BH growth rate would match the sSFR$=10^{-10}$
yr$^{-1}$ if the average SMBH growth rate is $t_{\rm Sal} \times {\rm
  sSFR} = 0.004$ or $0.4\%$ times the Eddington rate.  We consider the
time-averaged black hole specific accretion rate,

\begin{equation}
{\rm s}\dot{M}_{\rm BH} \equiv \frac{\dot{M}_{\rm BH}}{M_{\rm BH}}
\end{equation}

\noindent when exploring PZF parameters, $\lambda_{\rm Edd}$, $f_{\rm
  duty}$, and $t_{\rm AGN}$.  

\citet{sun15} also find general co-evolution of galaxies and their
black holes, which is consistent with no deviation in the $M_{\rm
  BH}/M_*$ ratio from $z=2\rightarrow0$.  However, they specifically
find that low-$z$ disk black holes appear to grow faster than their
galaxies, often with $\sdotMBH>10^{-9}$ yr$^{-1}$.  The time-averaged
$\lambda_{\rm Edd}$ in EAGLE simulations for a $10^7 \msolar$ SMBH at
$z=0-0.2$ is $0.01$ \citep{ros16}, which translates to
$\sdotMBH=2.5\times10^{-10}$ yr$^{-1}$ and is at least $2\times$
higher than the typical sSFR.  SMBH masses rapidly increase around
$M_{200}=10^{12} \msolar$ in EAGLE, which appears related to the
inability of gas heated by supernova-driven feedback to buoyantly rise
through the hot circumgalactic coronae that form at this halo mass
\citep{bow17,mca17}.  Inefficient feedback leads to rapid, non-linear
growth of the black hole.

We select our AGN models based on all these considerations: AGN
luminosities, duty cycles, lifetimes, and evolutionarily sustainable
SMBH growth rates.  Our models treat AGN as on-off light bulbs in this
exploration, even though a power law distribution for $\lambda_{\rm
  Edd}$ may be more appropriate given recent observational results
\citep[e.g.][]{jon16} and theoretical work \citep[e.g.][]{gab13}.  We
may explore such distributions in the future, but we do add a new AGN
flickering model based on the recent X-ray-derived accretion rate
distributions of \citet{air17}, who provide the average Eddington
ratio ($\lambda_{\rm Edd}\sim 0.1$) and duty cycle fraction ($1\%$)
above $\lambda_{\rm Edd}>0.01$.

\section{Fluctuating AGN ionizing the CGM of $L^*$ galaxies} \label{sec:results}

We choose our reference AGN model for the PZF effect: $L_{\rm
  bol}=10^{44.1} \ergs$ ($\lambda_{\rm Edd}= 0.1$), $f_{\rm
  duty}=10\%$, and $\tAGN=10^{6.2}$ yr.  This model, referred to as
{\it L441d10t6} using the nomenclature {\it L}[log($L_{\rm
    bol}/\ergs$)$\times10$]{\it d}[$f_{\rm duty}$ \%]{\it
  t}[log($t_{\rm AGN}$/yr)] and listed in bold in Table
\ref{tab:PZFmodels}, has a $\sdotMBH=2.5\times 10^{-10} {\rm yr}^{-1}$
matching the time-averaged SMBH growth rate in EAGLE.  Most models we
explore have lower $\sdotMBH$ because 1) we want to explore the
minimum requirements for PZFs, and 2) the required parameter choices
prefer lower $\sdotMBH$, because first we want duty cycle fractions to
be small so that most star-forming galaxies do not appear as AGN, and
second Eddington-limited accretion is rare at low-$z$.  Seg17
performed a similar parameter exploration, using parameters
$\lambda_{\rm Edd}$ (spanning $0.01-1$), $f_{\rm duty}$ (spanning
$1-50\%$), and $\tAGN$ (spanning $10^5-10^7$ yr plus considering
$10^3-10^5$ yr for a special case).  Seg17 also performed a wider
parameter exploration, considering multiple redshift ($z=3.0$, $0.1$)
and galaxy masses ($M_*=10^{10}$, $10^{11} \msolar$), whereas we are
applying PZFs to COS-Halos star-forming galaxies.

\begin{figure*}
\includegraphics[width=0.90\textwidth]{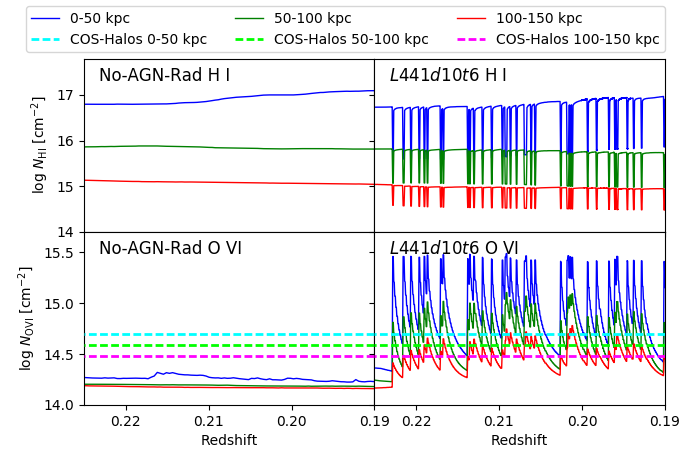}
\includegraphics[width=0.325\textwidth]{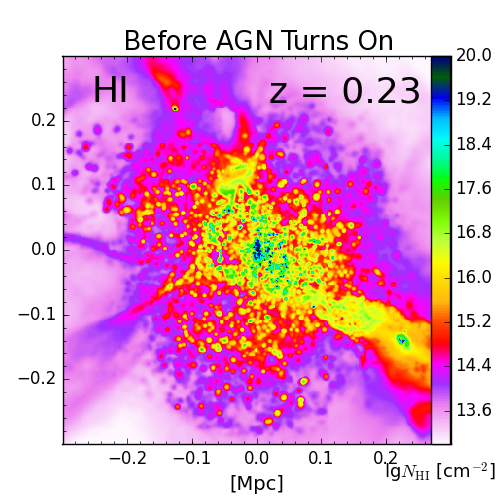}
\includegraphics[width=0.325\textwidth]{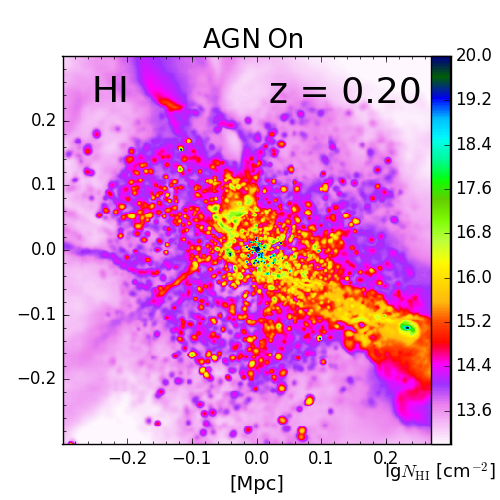}
\includegraphics[width=0.325\textwidth]{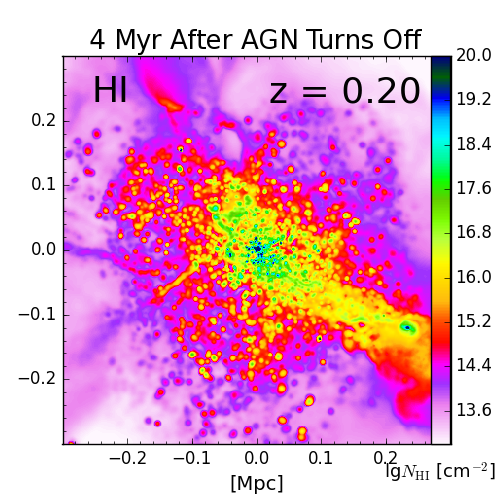}
\includegraphics[width=0.325\textwidth]{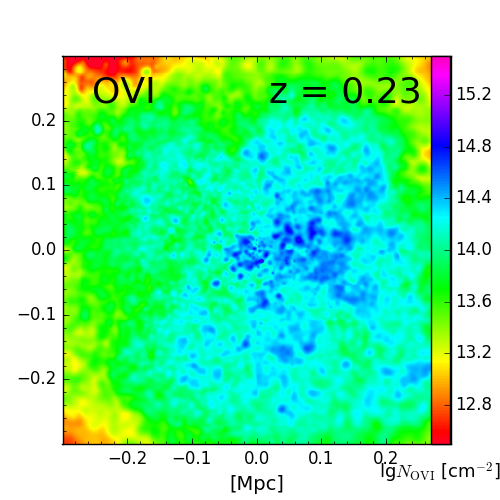}
\includegraphics[width=0.325\textwidth]{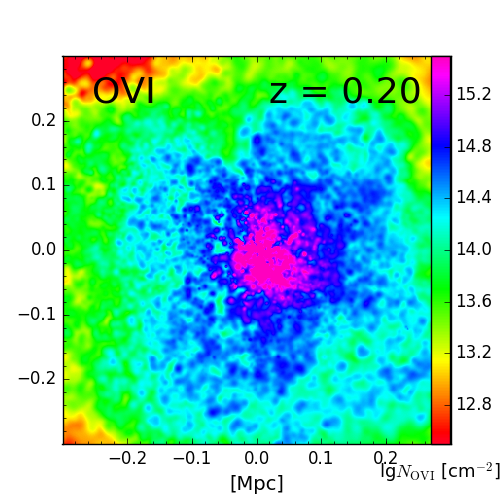}
\includegraphics[width=0.325\textwidth]{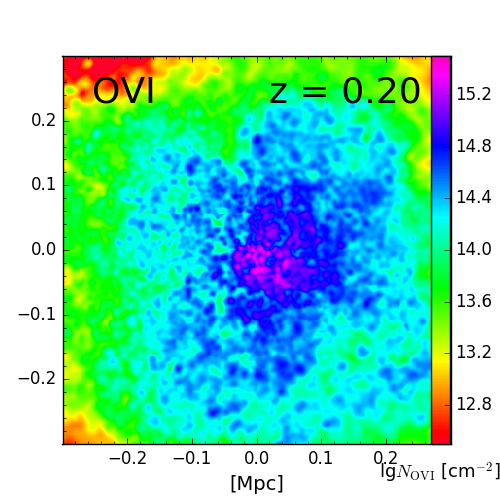}
\caption[]{Upper portion: Histories of median $\HI$ (upper panels) and
  $\OVI$ (lower panels) column densities in 3 impact parameter bins
  ($b=0-50$, $50-100$, \& $100-150$ kpc) for the No-AGN-Rad model
  (left) and the reference {\it L441d10t6} ($L_{\rm bol}=10^{44.1}
  \ergs$, $f_{\rm duty}=10\%$, and $\tAGN=10^{6.2}$ yr) PZF model
  (right).  COS-Halos $\OVI$ medians for the star-forming sample
  compiled from \citet{wer13} in the three impact parameter bins are
  shown as dashed lines for comparison; their typical $1-\sigma$
  dispersions are $\approx 0.2$ dex (not shown).  Lower Portion:
  $600\times 600$ kpc snapshots are shown for $\HI$ (upper panels) and
  $\OVI$ (lower panels) at three times: before the AGN turns on
  (left), when the AGN is on (middle), and 4 Myr after the AGN turns
  off (right).}
\label{fig:ts_snaps_fid}
\end{figure*}

\begin{table*}
\caption{Proximity zone fossil models}
\begin{tabular}{lrrrrrrrr}
\hline
Name & 
Resolution &
$L_{\rm bol}^a$ &
$f_{\rm duty}$ &
$\tAGN^b$ &
$\sdotMBH^c$ &
Isotropic? &
$N_{\OVI,0-75}^{d,e}$ &
$N_{\OVI,75-150}^{d,f}$ 
\\
\hline
\multicolumn {9}{c}{}\\
No-AGN-Rad & {\it M5.3} & -- & -- & -- & -- & -- & $14.19^{+0.15}_{-0.15}$ & $14.17^{+0.15}_{-0.17}$ \\
\textbf {\textit {L441d10t6}} & {\it M5.3} & 44.1 & 10 & 6.2 & $2.5\times 10^{-10}$ & yes & $14.62^{+0.42}_{-0.32}$ & $14.42^{+0.26}_{-0.23}$ \\
\textit{L441d10t7} & {\it M5.3} & 44.1 & 10 & \textbf{7.1} & $2.5\times 10^{-10}$ & yes & $14.36^{+0.38}_{-0.21}$ & $14.30^{+0.26}_{-0.21}$ \\
\textit{L441d10t5} & {\it M5.3} & 44.1 & $10$ & \textbf{5.0} & $2.5\times 10^{-10}$ & yes & $14.89^{+0.28}_{-0.30}$ & $14.47^{+0.20}_{-0.21}$ \\
\textit{L431d10t6} & {\it M5.3} & \textbf{43.1} & 10 & 6.2 & $2.5\times 10^{-11}$ & yes & $14.33^{+0.23}_{-0.19}$ & $14.20^{+0.15}_{-0.17}$ \\
\textit{L436d10t6} & {\it M5.3} & \textbf{43.6} & 10 & 6.2 & $8\times 10^{-11}$ & yes & $14.43^{+0.30}_{-0.24}$ & $14.29^{+0.16}_{-0.19}$ \\
\textit{L451d10t6} & {\it M5.3} & \textbf{45.1} & 10 & 6.2 & $2.5\times 10^{-9}$ & yes & $14.84^{+0.35}_{-0.38}$ & $14.67^{+0.38}_{-0.34}$ \\
\textit{L441d01t5} & {\it M5.3} & 44.1 & \textbf{1.0} & \textbf{5.0} & $2.5\times 10^{-11}$ & yes & $14.32^{+0.21}_{-0.16}$ & $14.23^{+0.14}_{-0.18}$ \\
\textit{L441d03t5} & {\it M5.3} & 44.1 & \textbf{3.2} & \textbf{5.0} & $8\times 10^{-11}$ & yes & $14.63^{+0.28}_{-0.27}$ & $14.32^{+0.15}_{-0.18}$ \\
\textit{L441d03t6} & {\it M5.3} & 44.1 & \textbf{3.2} & 6.2 & \textbf{$8\times 10^{-11}$} & yes & $14.35^{+0.37}_{-0.21}$ & $14.28^{+0.20}_{-0.20}$ \\
\textit{L431d32t6} & {\it M5.3} & \textbf{43.1} & \textbf{32} & 6.2 & \textbf{$8\times 10^{-11}$} & yes & $14.50^{+0.27}_{-0.20}$ & $14.30^{+0.16}_{-0.18}$ \\
\textit{L446d01t5} & {\it M5.3} & \textbf{44.6} & \textbf{1.0} & \textbf{5.0} & \textbf{$8\times 10^{-11}$} & yes & $14.62^{+0.47}_{-0.29}$ & $14.33^{+0.19}_{-0.20}$ \\
\textit{L436d10t5} & {\it M5.3} & \textbf{43.6} & 10 & \textbf{5.0} & \textbf{$8\times 10^{-11}$} & yes & $14.59^{+0.27}_{-0.24}$ & $14.30^{+0.17}_{-0.20}$ \\
\textit{L431d32t5} & {\it M5.3} & \textbf{43.1} & \textbf{32} & \textbf{5.0} & \textbf{$8\times 10^{-11}$} & yes & $14.64^{+0.25}_{-0.27}$ & $14.33^{+0.15}_{-0.19}$ \\
\textit{L441d10t6-bicone} & {\it M5.3} & 44.1 & 10 & 6.2 & $2.5\times 10^{-10}$ & {\bf biconical} & $14.45^{+0.36}_{-0.26}$ & $14.32^{+0.23}_{-0.21}$ \\
\textit{L441d10t5-bicone} & {\it M5.3} & 44.1 & 10 & \textbf{5.0} & $2.5\times 10^{-10}$ & {\bf biconical} & $14.66^{+0.35}_{-0.34}$ & $14.34^{+0.23}_{-0.23}$ \\
No-AGN-Rad-{\it M4.4}        & \textbf{\textit{ M4.4}} & -- & -- & -- & -- & -- & $14.33^{+0.16}_{-0.14}$ & $14.36^{+0.15}_{-0.15}$ \\ 
\textit{L441d10t6-M4.4} & \textbf{\textit{ M4.4}} & 44.1 & 10 & 6.2 & $2.5\times 10^{-10}$ & yes & $14.54^{+0.26}_{-0.20}$ & $14.48^{+0.17}_{-0.1}$ \\
\textit{L441d10t5-M4.4} & \textbf{\textit{ M4.4}} & 44.1 & 10 & \textbf{5.0} & $2.5\times 10^{-10}$ & yes & $14.75^{+0.32}_{-0.21}$ & $14.57^{+0.10}_{-0.19}$ \\
\hline
COS-Halos   & -- & --    & --     & --   & --    &  --               & $14.71^{+0.24}_{-0.15}$ & $14.52^{+0.14}_{-0.18}$ \\
\hline
\end{tabular} 
\\
\parbox{25cm}{
$^a$ log erg s$^{-1}$, $^b$ log yr, $^c$ yr$^{-1}$, $^d$ cm$^{-2}$, $^e$ $0-75$ kpc, $^f$ $75-150$ kpc
}
\label{tab:PZFmodels}
\end{table*}

We present our main results in this section, beginning with the PZF
effect assuming isotropic emission and varying the parameters $L_{\rm
  bol}$, $\tAGN$, and $f_{\rm duty}$.  We then extend our analyses to
models with anisotropic emission. 

\subsection{Isotropic models}

The evolution of $N_{\HI}$ and $N_{\OVI}$, in the No-AGN-Rad model and
our reference model, {\it L441d10t6}, are plotted in the upper set of
panels of Figure \ref{fig:ts_snaps_fid}.  We show $380$ Myr histories of the
median $\HI$ (upper panels) and $\OVI$ (middle panels) column
densities in 3 impact parameter bins ($b=0-50$, $50-100$, \& $100-150$
kpc).  The $\OVI$ column densities in the No-AGN-Rad model (left) are
too weak by a factor of $2-3$, showing little variation with time.
The {\it L441d10t6} panels (right) span $34$ AGN-on phases lasting
$1.5$ Myr each, which are indicated by a dip in $\HI$ and a spike in
$\OVI$.  The $\HI$ returns rapidly to its equilibrium ionization level
on a timescale of $\approx 10^5$ yr, which is too short to be seen on
this plot.  $\OVI$ on the other hand shows a prolonged decline for
every AGN-on phase, illustrating how the delayed recombination effect
occurs for metal ions but not $\HI$.

We show three sets of $\HI$ and $\OVI$ column density maps in the
lower portion of Fig. \ref{fig:ts_snaps_fid} to illustrate the general
behaviour of PZFs.  The column density maps are $600 \times 600$ kpc
across with 1 kpc pixel sizes, which we find to be converged with the
column densities being insensitive to pixel size at this size and
smaller.  The first snapshots at $z=0.23$ (left) show a normal CGM
unaffected by the AGN.  We display an AGN-on phase just below $z=0.20$
in the middle panels, where $\HI$ is diminished and $\OVI$ is
enhanced, especially in the inner 100 kpc.  The right panel shows the
CGM $4$ Myr after the AGN turns off, when $\HI$ has returned to its
earlier appearance, but $\OVI$ is still enhanced due to the delayed
recombination PZF effect.

To compare directly with the COS-Halos blue galaxy sample, we use the
python module called Simulation Mocker Of Hubble Absorption-Line
Observational Surveys (SMOHALOS) introduced by Opp16 to create mock
COS-Halos surveys using the observed impact parameters of that survey.
Unlike Opp16, who used 20 zooms at 6 different redshifts to find the
simulated galaxy that matched the observed galaxy most closely in
terms of $M_*$ and sSFR, here we use one evolving zoom at every
redshift output between $z=0.225$ and $0.15$ where the AGN is off,
which gives us $193$ out of $213$ snipshots.  Mock column densities
are selected from column density maps along the $x$, $y$, and $z$
projections with 1 kpc pixel resolution, which we tested for
resolution convergence at this pixel size and below.  Time evolving
visualizations of our column density maps are available at
http://noneq.strw.leidenuniv.nl/PZF/.

\begin{figure}
\includegraphics[width=0.49\textwidth]{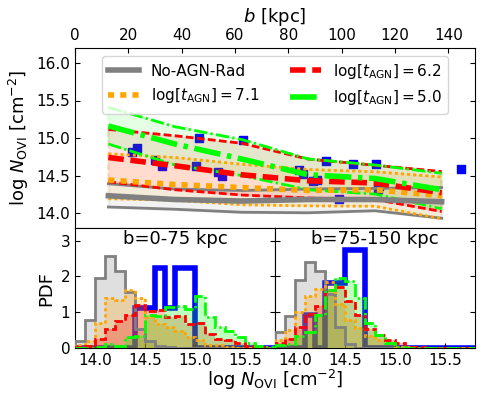}
\caption[]{{\it Upper panel:} Mock $\OVI$ SMOHALOS realizations of the
  COS-Halos star-forming $L^*$ sample comparing medians of the
  No-AGN-Rad model (solid grey line), the reference {\it L441d10t6}
  PZF model (dashed red line), and two other models with different AGN
  lifetimes $\tAGN=10^{7.1}$ yr ({\it L441d10t7}, orange dotted) and
  $10^{5.0}$ yr ({\it L441d10t5}, green dot-dashed).  One-$\sigma$
  dispersions are indicated by thin lines and the shaded region for
  each model.  COS-Halos data (blue squares) is plotted for
  comparison.  {\it Lower panels:} Histograms of the SMOHALOS
  realizations (filled histograms) divided into two impact parameter
  bins ($0-75$ \& $75-150$ kpc) compared to the COS-Halos observed
  histograms (thick blue histogram lines).  The PZF effect
  significantly enhances $\OVI$ column densities, and AGN with shorter
  lifetimes but the same time-averaged power have greater PZF
  effects. }
\label{fig:b_OVI_tAGN}
\end{figure}

In the upper panel of Figure \ref{fig:b_OVI_tAGN}, we plot the median
SMOHALOS $\OVI$ column densities in 6 bins between $0$ and $150$ kpc,
with $1$-$\sigma$ dispersions for the No-AGN-Rad model (grey) and the
reference model (red).  The No-AGN-Rad median is quite similar to the
Opp16 SMOHALOS run for the star-forming COS-Halos galaxies, but has a
smaller average $1-\sigma$ dispersion (cf. $0.15$ dex here, $0.3$ dex
in Opp16).  The main difference is that the Opp16 blue sample includes
5 additional $M_*>10^{10.5} \msolar$ star-forming galaxies, which have
lower $\OVI$ columns in both SMOHALOS and COS-Halos.  This contributes
to the greater Opp16 dispersion measure, as well as using ten galaxy
zooms to simulate a wider range of $M_*$ and SSFR in Opp16 versus
using one galaxy here.

The COS-Halos $\OVI$ column densities have a dispersion of $\approx
0.15-0.20$ dex as well, but have much higher observed columns
(log[$N_{\OVI}$/cm$^{-2}$]$=14.7$, $14.6$, \& $14.5$ for $b=0-50$,
$50-100$, \& $100-150$ kpc, respectively), which is $3\times$ higher
than our No-AGN-Rad model with $N_{\OVI}=10^{14.2} \cms$ inside 50 kpc
and and $2\times$ higher at beyond 100 kpc, in agreement with Opp16.
The reference PZF model shows an increase over the No-AGN-Rad model at
all impact parameters, from 0.43 dex for $b=0-75$ kpc to 0.25 dex for
$b=75-150$ kpc, resulting in column densities only $0.1$ dex below
COS-Halos.  Unsurprisingly, the PZF increases the $1-\sigma$
dispersions from $0.15$ dex to $0.25-0.4$ dex.

\subsubsection{Parameter variation}

\begin{figure*}
\includegraphics[width=0.90\textwidth]{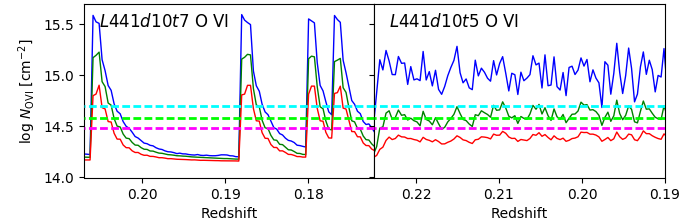}
\caption[]{$\OVI$ time histories are displayed for the {\it L441d10t7}
  ($\tAGN=10^{7.1}$ yr, left) and {\it L441d10t5} ($\tAGN=10^{5.0}$
  yr, right) models as in Fig. \ref{fig:ts_snaps_fid}.  The same
  time-averaged power results in very different $\OVI$ column
  densities with the left panel showing 4 distinct AGN-on episodes and
  the right panel achieving a near steady ionization state with over
  300 AGN-on episodes.}
\label{fig:ts_tAGN}
\end{figure*}

\noindent{\bf AGN lifetime:} Figure \ref{fig:b_OVI_tAGN} (lower
panels) also shows SMOHALOS realizations for models {\it L441d10t7}
and {\it L441d10t5} corresponding to longer and shorter AGN lifetimes,
respectively (but the same luminosity and duty cycle).  Despite
delivering the same time-averaged AGN radiation power, the PZF effect
is stronger for shorter lifetimes, which pump the CGM ionization level
more frequently while the recombination timescale remains the same.
OS13 and Seg17 also showed this non-linear PZF ionization effect,
although if the ionization timescale is longer than $\tAGN$, the
maximum ionization state will not be achieved.  The evolution of {\it
  L441d10t7} and {\it L441d10t5} in Figure \ref{fig:ts_tAGN} bear out
this behaviour-- $\OVI$ reaches higher columns for the $\tAGN=
10^{7.1}$ yr during the AGN-on phase, while $\tAGN= 10^{5.0}$ yr
achieves lower $\OVI$ columns, but far more frequently and thereby
reaching higher average $N_{\OVI}$.

The SMOHALOS $\OVI$ columns during the $\approx 90\%$ of the galaxy's
history with inactive AGN better match the observation for the {\it
  L441d10t5} model than for the reference {\it L441d10t6} model.  The
histograms in Fig. \ref{fig:b_OVI_tAGN} show an excellent fit to the
$b=75-150$ kpc COS-Halos $\OVI$ distribution, and over-predict the
inner column densities by $\approx 0.2$ dex.  With over 300 AGN-on
phases contributing to the time history shown in Figure
\ref{fig:ts_tAGN} for {\it L441d10t5} versus only four for {\it
  L441d10t7}, the ``flickering'' $\tAGN=10^{5.0}$ yr model has
achieved a relatively steady ionization state where the $\OVI$ level
has little correlation with AGN activity.  {\it This model shows that
  for normal expectations for SMBH parameters, $\OVI$ columns
  comparable to and even in excess of observed COS-Halos $\OVI$ are
  plausible.}

\begin{figure}
\includegraphics[width=0.49\textwidth]{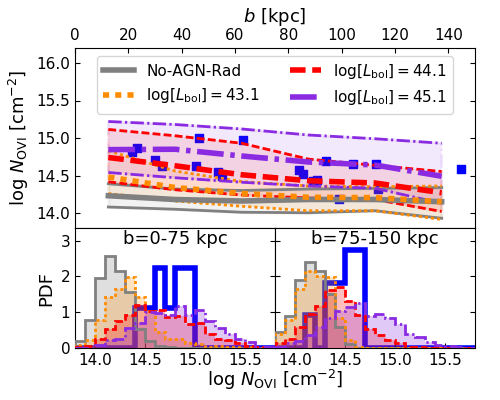}
\caption[]{The dependence of $\OVI$ column density profiles on AGN
  luminosity for the SMOHALOS realizations are plotted as in
  Fig. \ref{fig:b_OVI_tAGN}.  $\OVI$ column densities increase for
  stronger AGN, because more CGM oxygen in lower ions is ionized up to
  $\OVI$.  The Eddington limited $L_{\rm bol}=10^{45.1} \ergs$ AGN
  creates too much $\OVI$ with too large of a dispersion compared to
  COS-Halos in the lower panels, while the $L_{\rm bol}=10^{43.1}
  \ergs$ AGN is too weak.}
\label{fig:b_OVI_Lbol}
\end{figure}

\noindent{\bf AGN luminosity:} $N_{\OVI}$ increases with AGN
luminosity, as shown in Fig. \ref{fig:b_OVI_Lbol} using $f_{\rm duty}
= 10\%$ and $\tAGN = 10^{6.2}$ yr.  OS13 showed that $\OVI$ columns
can decline in PZFs if the AGN is strong and the density is low
because more $\OVI$ becomes ionized to $\OVII$ and above.  However,
the CGM is too dense and the AGN explored here are too weak to reach
this limit.  The best fit to the data is still the $L_{\rm bol} =
10^{44.1} \ergs$ ($\lambda_{\rm Edd} = 0.1$) model, and this
exploration shows that low luminosity AGN ($L_{\rm bol}\sim
10^{43.1}\ergs$) are too weak to ionize $\OVI$ to the observed levels
using the \citet{saz04} spectrum, at least for $f_{\rm duty}=10\%$ and
$\tAGN=10^{6.2}$ yr.  The $L_{\rm bol} = 10^{45.1} \ergs$ model
predicts too much $\OVI$ while implying a likely unsustainable
$\sdotMBH=2.5\times 0^{-9}$ yr$^{-1}$ for $M_{\rm BH} \sim 10^{7}
\msolar$.

\begin{figure}
\includegraphics[width=0.49\textwidth]{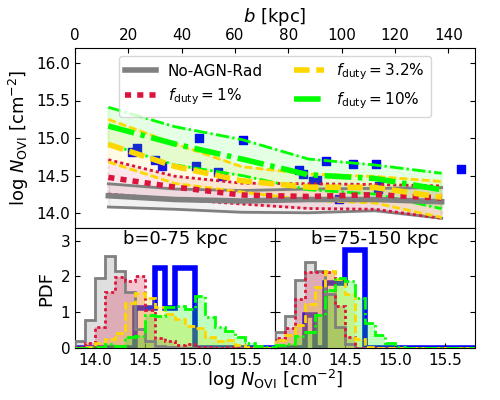}
\caption[]{The dependence of $\OVI$ column density profiles on duty
  cycle fraction for the SMOHALOS realizations are plotted as in
  Fig. \ref{fig:b_OVI_tAGN}.  Using $\tAGN=10^5$ yr and $L_{\rm
    bol}=10^{44.1} \ergs$, the $\OVI$ columns decrease for lower duty
  cycles, owing to declining AGN time-averaged power.}
\label{fig:b_OVI_d}
\end{figure}

\noindent{\bf Duty cycle:} We show the dependence of COS-Halos $\OVI$
on the duty cycle in Figure \ref{fig:b_OVI_d} for $\tAGN=10^{5.0}$ yr
and $L_{\rm bol} = 10^{44.1} \ergs$.  The PZF effect significantly
declines at lower duty cycle, even though the intervals in between AGN
activity average $10^7$ yr for $f_{\rm duty}=1\%$, which is
approximately the recombination timescale for $\OVI$.  This contrasts
to cases in OS13 where the PZF effect did not show much dependence on
the duty cycle as long as the interval time was shorter than the ion
recombination time.  The reason for the difference here is that for
the short AGN lifetimes the CGM does not reach ionization equilibrium
with the enhanced AGN+EGB field .  In the limit of short AGN lifetimes,
Seg17 demonstrated the PZF approaches a quasi-steady-state ionization
level described by the time averaged AGN+EGB field.  Here we see the
effect of the flickering AGN with higher duty cycles, and therefore
higher time-averaged power, reaching higher ionization levels.

\begin{figure}
\includegraphics[width=0.49\textwidth]{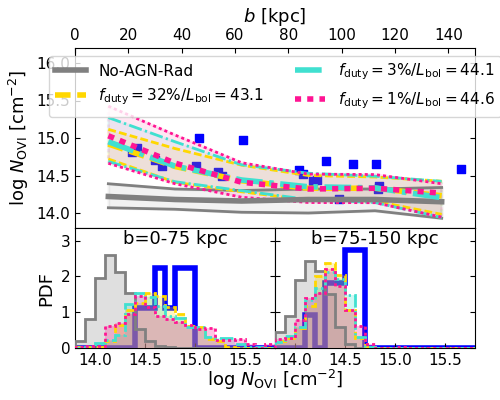}
\caption[]{SMOHALOS realizations are plotted as in
  Fig. \ref{fig:b_OVI_tAGN} using $\tAGN=10^5$ yr, but keeping
  $\sdotMBH=8\times 10^{-11}$ yr$^{-1}$ through a combination of
  varying duty cycle and AGN luminosity.  There is little dependence
  on how AGN radiation is distributed in time given this AGN average
  power.  The $f_{\rm duty}=10\%$, $L_{\rm bol}=10^{43.6} \ergs$ model
  ({\it L436d10t5}, not shown) shows identical statistics to the other
  PZF models shown here. }
\label{fig:b_OVI_sMdot}
\end{figure}

\noindent{\bf Time-averaged SMBH growth rate kept constant:} The final
parameter variation we explore in isotropic models is leaving
$\sdotMBH$ constant at $8\times10^{-11}$ yr$^{-1}$ but varying $f_{\rm
  duty}$ and $L_{\rm bol}$ as we show in Figure \ref{fig:b_OVI_sMdot}
for $\tAGN=10^{5.0}$ yr.  Our hypothesis is that there may exist a
combination of duty cycle and AGN luminosity that maximizes the PZF
effect while keeping the SMBH growth rate similar to the sSFR
($\approx 10^{-10}$ yr$^{-1}$).  Despite a factor of $32$ difference
in AGN luminosity and duty cycle explored, there is surprsingly little
difference in the PZF effect, especially for the $\tAGN=10^{5.0}$ yr
lifetime.  Models with $\tAGN=10^{6.2}$ yr appear in Table
\ref{tab:PZFmodels}, and show the high duty cycle fraction, $f_{\rm
  duty}=32\%$ model having slighly higher inner $\OVI$ column
densities, but otherwise very little difference.  Higher $f_{\rm
  duty}$ values are disfavoured, since more galaxies would appear as
AGN.

The exploration here shows that the PZF effect does not depend
strongly on how the AGN radiative power is distributed in time for
$\sdotMBH= 8\times 10^{-11}$ yr$^{-1}$, which corresponds to a
time-averaged AGN luminosity of $\langle L_{\rm bol}\rangle =
10^{42.6} \ergs$.  Figure 10 of Seg17 explicitly demonstrates this
effect.  In the limit of AGN interval times $\ll t_{\rm rec}$, the column
densities approach the steady state of constant ionization by the time
averaged AGN+EGB field.

Interestingly, the {\it L446d01t5}, which is plotted in
Fig. \ref{fig:b_OVI_sMdot} in dotted pink, is a new
model\footnote{Added in the revision as a suggestion from the
  referee.}  that matches the averaged properties of an AGN occupying
$M_*=10^{10}-10^{10.5} \msolar$ star-forming galaxies according to
\citet{air17}.  Although, they derive distributions of $\lambda_{\rm
  Edd}$ as a function of galaxy redshift, mass, and type (star-forming
vs. quiescent), their Figure 6 gives the mean duty cycle fraction
above $\lambda_{\rm Edd}>0.01$ ($1\%$) and their Figure 8 gives the
average Eddington ratio above $\lambda_{\rm Edd}>0.01$ ($\lambda_{\rm
  Edd}\sim 0.1$).  The $L_{\rm bol}=10^{44.6} \ergs$ luminosity
implies $\lambda_{\rm Edd}=0.3$ for our assumed $M_{\rm BH}=10^{7.0}
\msolar$, but because \citet{air17} base their SMBH masses using
$M_{\rm BH}=0.002 M_{*}$, their model implies $M_{\rm BH}\approx
10^{7.5} \msolar$.  Nevertheless, this demonstrates a significant PZF
effect using the population characteristics from a recent
X-ray-selected survey assuming a randomly flickering SMBH with average
interval times between AGN episodes of $\approx 10^{7}$ yr.


\subsection{Anisotropic models}

Thus far we have explored isotropically emitting models, even though
optically thick dust tori are expected to surround the central SMBH
engine during accretion episodes.  Here we explore a simple
anisotropic AGN model by assuming a biconical opening angle of
$120^{\circ}$ corresponding to a covering factor of $2\pi$ steradians.
We set the axis of the bicone using the angular momentum vector of
star-forming gas within $3\hkpc$ at that timestep, which serves as a
proxy for shielding by dust in the torus and/or extended dust in the
galactic disk.  No AGN radiation leaks in the equatorial direction
corresponding to the other $2\pi$ steradians, which is an assumption
that may underestimate the PZF effect if the dust torus is not Compton
thick, as we discuss in \S\ref{sec:discuss}.  We also run a {\it
  L441d10t6-bicone} model with the bicone aligned to the stellar disk
within $10\hkpc$, and find that the cone axis varied less, but there
is hardly any difference in the $\OVI$ statistics.  We also allow our
bicone to have random orientations, which applies the assumption that
the black hole accretion disk is randomly oriented with respect to the
galactic disk and is responsible for obscuration, and find no
difference in the $\OVI$ statistics.  

\begin{figure}
\includegraphics[width=0.49\textwidth]{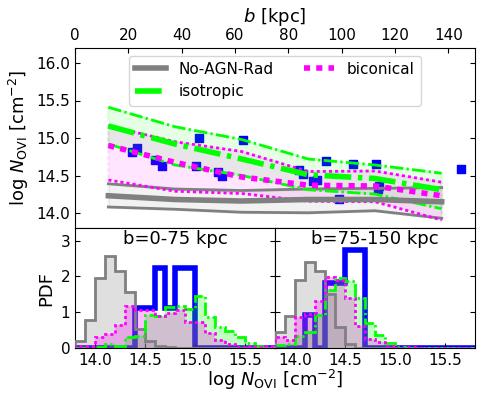}
\caption[]{SMOHALOS realizations are plotted as in
  Fig. \ref{fig:b_OVI_tAGN} using $\tAGN=10^5$ yr models, but showing
  the effect of using a biconical opening angle ({\it
    L441d10t5-bicone}) compared to the isotropic model ({\it
    L441d10t5}).  The AGN radiative power is half as strong in the
  latter, but still creates a significant PZF effect.}
\label{fig:b_OVI_theta}
\end{figure}

The SMOHALOS results are shown in Figure \ref{fig:b_OVI_theta},
showing that the $\OVI$ column densities are intermediate between the
No-AGN-Rad model and the isotropic cases.  While this is not
surprising, the biconical $\OVI$ median column density is stronger
than halfway between the No-AGN-Rad and isotropic cases even though
half the CGM volume is ionized with a $120^{\circ}$ opening angle.
This is particularly apparent for the $\tAGN=10^{5.0}$ yr biconical
model ({\it L441d10t5-bicone}, magenta dotted line).  Part of the
reason for this behaviour is that more than half of the sight lines
are enhanced due to the galaxy-line of sight geometry-- even if only
half the volume is ionized, more than half the sight lines intersect
the ionization cone, even for a $90^{\circ}$ inclined torus.  We
compile statistics in the $x$, $y$, and $z$ projections regardless of
torus orientation, because the torus orientation changes and the three
projection axes should average out.

If the dust obscuration primarily arises from the BH accretion disk
that is decoupled from the galactic disk orientation, then a model
with random orientation for each AGN episode is more appropriate.  We
therefore run a {\it L441d10t5-bicone} model (not shown) where the
bicone axis orientation is randomly oriented, which yields identical
medians and dispersions as the gas-aligned bicone case.  Our AGN
bicone axis orientation does not appear to affect the statistics of a
COS-Halos-like survey.  Short AGN lifetimes again achieve a
quasi-steady-state ionization level with the AGN+EGB field, where the
AGN field is half as strong as in the isotropic case.

The {\it L441d10t5-bicone} model provides a very good fit to the
COS-Halos data.  Some of this model's column densities at larger $b$
are slightly low compared to COS-Halos, but all the highest data
points are within the $2-\sigma$ range (not shown).  It is not
worthwhile to tweak the model to get a better match.  We can certainly
find a better matching model with a slightly larger opening angle, a
slightly larger duty cycle or AGN luminosity, a shorter AGN lifetime,
or some other combination of all of these.  However, we have shown
that this approximate model, which grows its SMBH at an evolutionarily
sustainable rate, satisfies the COS-Halos $\OVI$ statistics.  For
consistency with other SMOHALOS models that apply statistics only to
phases when the AGN is off and the galaxy appears to be a normal
star-forming galaxy, we should include snipshot frames with active but
obscured AGN.  However, we do not do so because this makes only a
small difference.  More importantly, this means that 5\%, instead of
10\% of star-forming galaxies will be observed as active AGN in the
optical, better reconciling the non-AGN COS-Halos detections on the
BPT diagram \citep{wer12}.

\subsection{Other Ions} \label{sec:lowions}

\begin{figure*}
\includegraphics[width=0.49\textwidth]{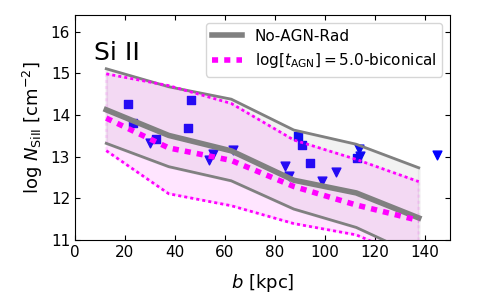}
\includegraphics[width=0.49\textwidth]{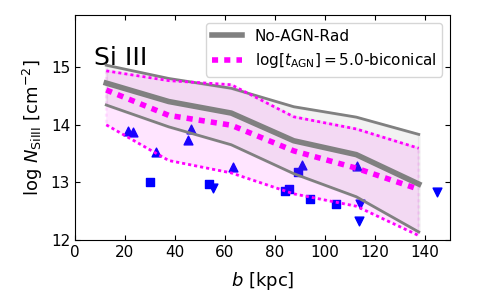}
\includegraphics[width=0.49\textwidth]{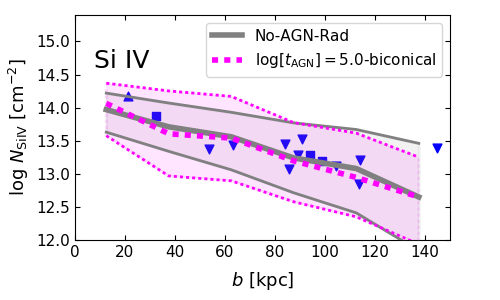}
\includegraphics[width=0.49\textwidth]{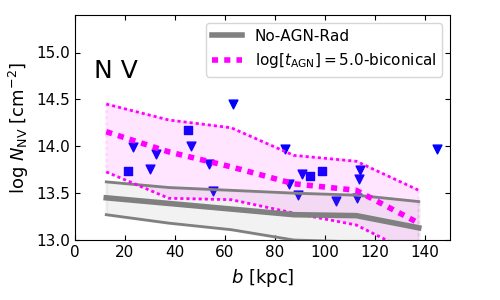}
\caption[]{Other ions are displayed for the {\it L441d10t5-bicone}
  model (dotted magenta) relative to the No-AGN-Rad model (solid
  grey), including $\SiII$ (upper left), $\SiIII$ (upper right),
  $\SiIV$ (lower left), and $\NV$ (lower right).  Medians (thick
  lines) and $1$-$\sigma$ dispersion (thin lines) are displayed.  Blue
  squares indicate detections, upside-down triangles are upper limits
  for non-detections, and right-side-up triangles are lower limits for
  saturated lines.  Low ions ($\SiII$, $\SiIII$) are moderately
  reduced by the PZF effect, high ions ($\OVI$, $\NV$, $\CIV$) are
  increased, and an intermediate ion like $\SiIV$ remain relatively
  unchanged.}
\label{fig:b_otherions}
\end{figure*}

Other metal ions should be affected by PZFs, including $\CIV$, which,
as Seg17 shows, can both be enhanced and reduced.  We consider low and
intermediate ions commonly observed around COS-Halos galaxies
($\SiII$, $\SiIII$, $\SiIV$), as well as $\NV$ for our favoured {\it
  L441d10t5-bicone} model in Figure \ref{fig:b_otherions}.  The lower
the ionization potential, the higher the physical density traced
\citep[e.g.][]{for13}, and therefore the shorter the typical
recombination time.  Low ions column densities are primarily reduced
in PZFs, while $\NV$, another high ion, is enhanced.  The intermediate
ion $\SiIV$ is at the threshold where it is neither reduced nor
enhanced significantly.  The $\SiII$ and $\SiIII$ medians are reduced by
$0.1-0.2$ dex, but the larger effect is an increased dispersion at
low column densities for these species.

Oppenheimer et al. (in prep.) compare SMOHALOS low-ion COS-Halos
measurements without considering PZFs, finding relatively good
agreement for Si species, although they note that $\SiIII$ is
$2-3\times$ too strong, which PZFs appear to help reduce.  That work
also finds good agreement between simulated and observed $\SiII$.
However, this ion is also sensitive to self-shielding, which enhances
$\SiII$ column densities by a factor of $\approx 2$ over the standard
uniform \citet{haa01} EGB model.  Combining PZFs with self-shielding
in simulations will assist the assessment of how these two effects
combine to alter low ion column densities.

$\NV$ is strongly enhanced, similar to $\OVI$, and appears to be
consistent with most COS-Halos observations.  \citet{wer16} explored
the $\NV/\OVI$ ratios in COS-Halos, showing most models indicate
tension with the low $\NV$ columns observationally ascertained mainly
using upper limit non-detections (plotted in the lower right panel of
Fig. \ref{fig:b_otherions}).  The PZF effect here does not appear to
alter $\NV/\OVI$ ratios for our one halo.  We do not show $\CIV$ since
it is not observed by COS-Halos, but this ion is enhanced by $\approx
0.2$ dex over the inner $125$ kpc as expected for an ion intermediate
between $\SiIV$ and $\NV$, as also shown by Seg17.

\begin{figure*}
\includegraphics[width=0.49\textwidth]{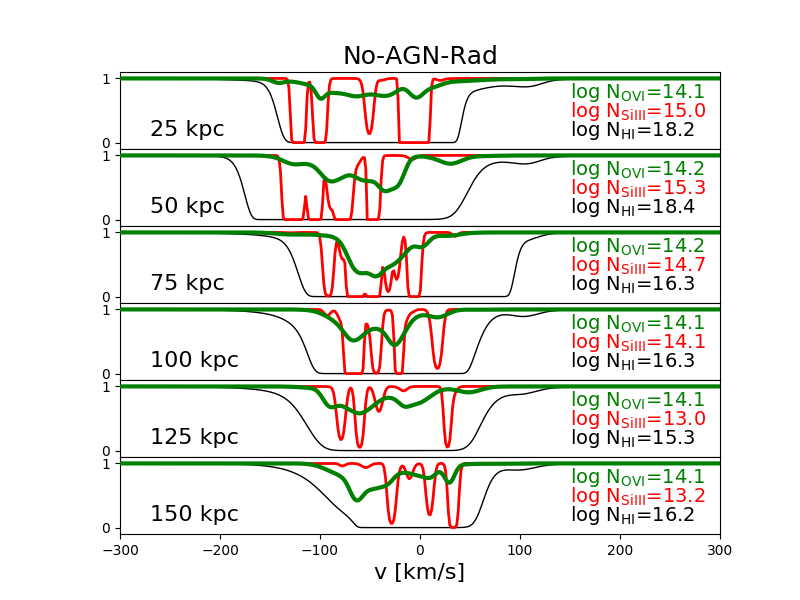}
\includegraphics[width=0.49\textwidth]{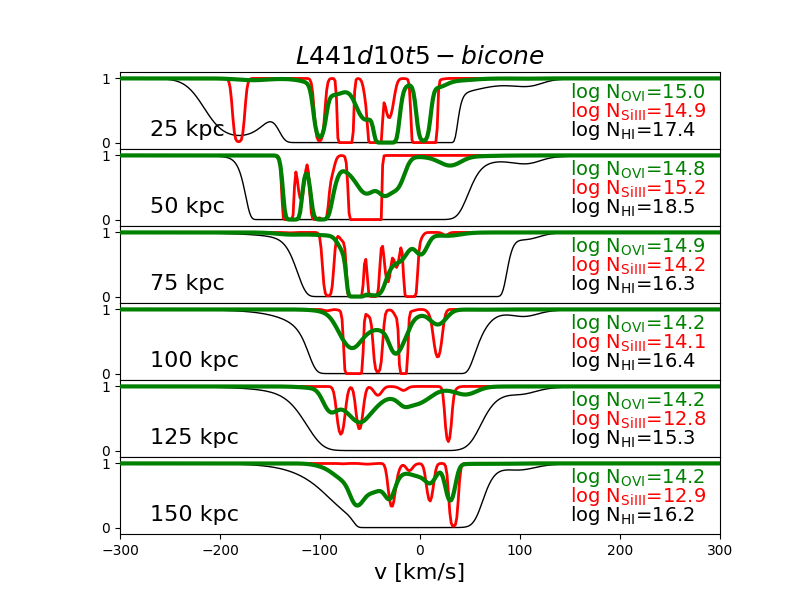}
\caption[]{Mock spectra generated using SpecWizard under idealized
  conditions (no noise or instrumental broadening added).  Six LOS
  spanning $b=25-150$ kpc show $\OVI$ (thick green), $\SiIII$ (red),
  and $\HI$ (thin black) for the same sight lines in two $z=0.204$
  snipshots (No-AGN-Rad on left, \& {\it L441d10t5-bicone} PZF on
  right).  Total column densities integrated using SpecWizard are
  listed in each panel showing PZF increases in $\OVI$, slight
  decreases in $\SiIII$, and similar structures in $\HI$.  Note that
  these are two different simulation runs, which can lead to random
  differences that are not directly caused by the radiation field.}
\label{fig:spec}
\end{figure*}

Finally, we display idealized mock spectra in Figure \ref{fig:spec} at
6 impact parameters between 25 and 150 kpc.  SpecWizard, described in
\citet{the98} and \citet{sch03}, is used to generate idealized,
noiseless spectra without instrumental broadening.  The No-AGN-Rad
model (left) is compared to the {\it L441d10t5-bicone} when the AGN is
off (right) for three species ($\OVI$- green, $\SiIII$- red, and
$\HI$- black) in representative lines of sight (LOS) at $z=0.204$.
The same LOS are used in the two models from the same snipshot output,
and while the two models correspond to separate simulation runs since
$z=0.235$ resulting in slightly different gas distributions, the same
absorption line structures can be seen in almost every LOS as best
exemplified by $\HI$ at $50-150$ kpc.  As expected, $\OVI$ is
enhanced, especially at $b<100$ kpc, showing shallower smooth
components in the No-AGN-Rad model and narrower components in the PZF
model.  $\SiIII$ shows individual narrow component substructure in
both cases, which is slightly reduced in the PZF model.  Total
integrated column densities listed in each subpanel can be compared
across the runs, and the column densities are typical compared to the
column density ranges in Figs. \ref{fig:b_OVI_theta} ($\OVI$) and
\ref{fig:b_otherions} ($\SiIII$).

\citet{wer16} explored the component substructure using a similar plot
(their Fig. 5) and defined $\OVI$ components relative to their
alignment with lower metal ions, represented here by $\SiIII$.  They
find that about about 80\% of $\OVI$ absorbers are well-aligned with
low ions, with half being broad $\OVI$ absorbers ($b$-parameter $>40$
$\kms$) and half being narrow $\OVI$ absorbers ($<40$ $\kms$).  The
aligned $\SiIII$ lines are mostly narrow.  Our mock PZF spectra show
many more aligned $\OVI$-$\SiIII$ absorbers than the No-AGN-Rad model,
especially for narrower $\OVI$ components (e.g. $-100$ \& $0$ $\kms$
at $25$ kpc, $-130$ \& $-100$ $\kms$ at $50$ kpc, $30$ $\kms$ at $150$
kpc).  This is an encouraging finding given the \citet{wer16} results,
but a direct comparison between mock spectra and COS-Halos, where
instrumental broadening from the COS line spread function blends
together components and noise is included, is beyond the scope of this
work.  OS13 argued that PZF models can create aligned low and high
metal ion component absorbers as observed \citep[e.g.][]{tri11} that
are not possible with equilibrium models.  A key result is that our
narrower $\OVI$ components are related to the PZF while broader
components in the No-AGN-Rad model are related to the collisionally
ionized $\OVI$ (Opp16).  The latter still exist in our PZF models, but
the PZF adds narrow $\OVI$ mainly at smaller impact parameters.
Future work will further analyze mock spectra and compare to the
aligned absorber component categories introduced by \citet{wer16}.

\section{Discussion} \label{sec:discuss}

OS13 explored how flickering low-$z$ Seyfert galaxies might ionize
their local CGM, arguing that $\OVI$ column densities could be
significantly enhanced by photo-ionization.  Since that publication,
two main findings about AGN have strengthened the case for PZFs
existing around normal star-forming galaxies-- 1) shorter AGN
lifetimes (\S\ref{sec:des_lifetime}) and 2) SMBH growth rates similar
to, or in excess of, the stellar mass growth rate
(\S\ref{sec:des_mdot}).  {\it Our main result is that if a
  star-forming, disk galaxy grows its SMBH at the cosmologically
  expected rate and had luminous AGN activity within the previous
  $10-20$ Myr, then its CGM is highly likely to show enhanced $\OVI$
  due to the proximity zone fossil effect.}

Our exploration here represents a proof of concept that the highly
ionized CGM might be driven far out of ionization equilibrium by AGN
radiation in the low redshift Universe.  More sophisticated, follow-up
modelling needs to consider physically-derived SMBH accretion
histories \citep[e.g.][]{nov11, gab13}, radiative transfer with light
travel effects, as well as the non-equilibrium ionization from a
fluctuating field introduced here.  We expect light travel time
effects (included in Seg17) to be important when considering how the
CGM ionization level correlates with the AGN activity.  The light
travel time is $\approx 5\times 10^5$ yr out to 150 kpc, which is
significant compared to plausible AGN lifetimes.  The case of the
flickering AGN with $\tAGN\sim 10^5$ yr and $f_{\rm duty}\sim 10\%$
results in ionization that decorrelates from AGN activity, and
achieves what can be described as a PZF steady state.

Our obscured torus bicone model is simplistic, intended only to
explore the possibility of anisotropic AGN emission, which is
especially likely in low-luminosity AGN residing in disk galaxies with
significant dust.  Anisotropic models open the possibility of PZFs
arising and being enhanced due to dust obscuration that can change on
a short timescale, especially if the dust is associated with the SMBH
engine where its orbital time may be shorter than the AGN lifetime. It
is also unlikely that these AGN are Compton thick, meaning that X-ray
radiation will escape, and how such a spectrum ionizes the CGM is
beyond the scope of this paper.  Another crude approximation is the
binary nature of our AGN (either on or off), when in reality the AGN
emission spectrum escaping into the CGM depends on the amount and type
of obscuration.  Finally, we do not discount the possibility of
extreme UV or soft X-ray radiation from star formation escaping the
galaxy and ionizing $\OVI$ \citep[e.g.][]{can10, vas15}, possibly
anisotropically along the semi-minor axis of the galaxy.

The {\it L441d10t5-bicone} model represents our favoured model,
because 1) it agrees with the $\sdotMBH$ of an SMBH in a star-forming
disk, 2) the 10\% duty cycle (observed to be 5\% with obscuration) is
statistically reconcilable with observing no AGN in COS-Halos, 3) it
appeals to the shorter AGN lifetimes supported by recent work
\citep{gab13,schaw15}, and 4) its biconical ionization considers
obscuration by a dust torus and/or dust in the star-forming disk.  If
the dust obscuration is aligned with the stellar disk, then this could
help explain why \citet{kac15} observes that the azimuthal dependence
of $\OVI$ around bluer galaxies peaks along the semi-minor axis.  This
study partially motivated our consideration of the bicone model, since
\citet{kac15} suggest a wide opening angle of stronger $\OVI$ at $b\la
100$ kpc.  Our model provides an alternative to the model of star
formation-driven winds being responsible for strong $\OVI$ along the
semi-minor axis, and represents an addendum to the Opp16 model of
$\OVI$ ``haloes'' tracing the virialized $3\times 10^5$ K gas of $L^*$
galaxies with little relation to recent star formation.

{\bf Passive galaxies:} We do not explore the passive red COS-Halos
sample here, but the PZF effect could also enhance the circumgalactic
$\OVI$ columns for these galaxies.  Opp16 showed that while the
standard EAGLE zooms were consistent with most of the $\OVI$ upper
limits, three of the 12 COS-Halos passive galaxies showed
$N_{\OVI}=10^{14.2}-10^{14.4} \cms$ that were not reproduced in the
Opp16 SMOHALOS realizations.  There may be reason to believe the PZF
effect is not as common around these galaxies, because passive
galaxies are much less likely to be accreting the cold gas that feeds
AGN activity as well as star formation.  While one may expect less low
ionization cold gas in a passive galaxy's CGM, \citet{wer13} showed
that low and intermediate ions are nearly as common around passive
galaxies as around star-forming galaxies.  The predicted Opp16 $\OVI$
column densities for passive galaxies are low enough ($N_{\OVI}\sim
10^{13-13.5} \cms$) that the PZF effect could raise their values and
still be consistent with the COS-Halos measurements.

\begin{figure*}
\includegraphics[width=0.49\textwidth]{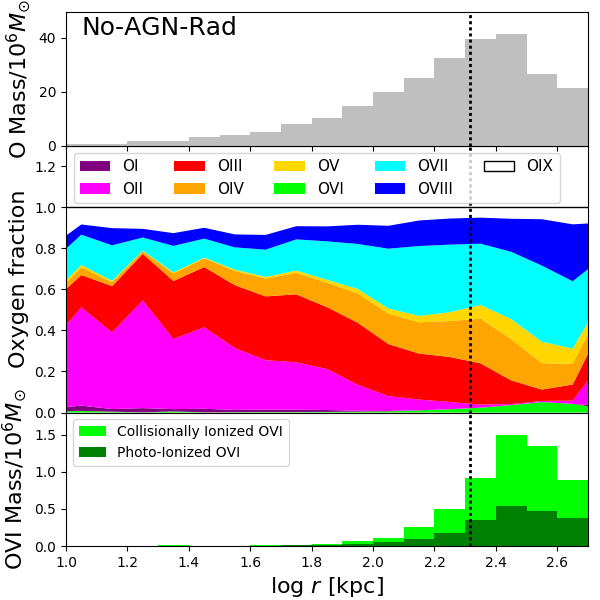}
\includegraphics[width=0.49\textwidth]{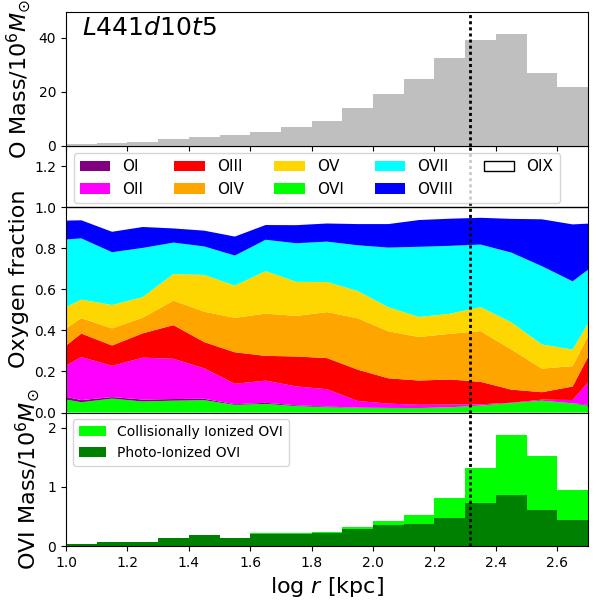}
\caption[]{The oxygen mass budgets and ionization fractions at
  $z=0.19$ plotted as functions of galactocentric radius for the
  No-AGN-Rad model (left) and the {\it L441d10t5} model (right).  The
  upper panels show oxygen mass, which is essentially identical in the
  two cases.  The center panels indicate the oxygen ionization
  fractions, and the lower panels show the $\OVI$ mass divided into
  collisionally ionized and photo-ionized components.  The dotted
  vertical line indicates the virial radius.  AGN photo-ionize $\OVI$
  mainly inside $R_{200}$ while $\OVI$ remains collisionally ionized
  mainly outside $R_{200}$.  Note that radii $\ga R_{200}$ will often
  be projected onto impact parameter $< R_{200}$.}
\label{fig:Omassbudget}
\end{figure*}

{\bf Collisionally ionized $\OVI$:} Opp16 argued that COS-Halos $\OVI$
is collisionally ionized tracing $T\sim 10^{5.5}$ K virialized gas in
and around $L^*$ haloes hosting normal star-forming galaxies, but they
did not consider radiation from flickering AGN.  In Figure
\ref{fig:Omassbudget} we plot the oxygen mass and ion fractions as a
function of radius from the galaxy for the {\it M5.3} No-AGN-Rad
(left) and a typical output for the {\it L441d10t5} PZF model (right)
at $z=0.19$ with the AGN off for approximately $0.5$ Myr.  The lower
panels show the $\OVI$ component divided into collisionally ionized
and photo-ionized components using a cut of $T=10^5$ K.  The $\OVI$
collisionally ionized component, mainly between $1-2 R_{200}$ (200-500
kpc; Opp16) remains for the PZF, but the PZF has an additional
photo-ionized $\OVI$ component, mainly inside $R_{200}$, that leads to
stronger $\OVI$.  Mock spectra in Fig. \ref{fig:spec} show narrow
components due to the PZF added to broader components from collisional
ionization.  The overall circumgalactic $\OVI$ ionization fraction of
gas within 500 kpc is $\approx 2\%$ in the No-AGN-Rad model (in
agreement with Opp16), while this fraction increases to $\approx
3-3.5\%$ for the {\it L441d10t5} PZF.  {\it The Opp16 model where
  $\OVI$ traces the virial temperatures of star-forming haloes still
  applies, but the PZF effect adds a comparable amount of
  photo-ionized $\OVI$ within the virial radius.}

{\bf IGM $\OVI$ statistics:} As mentioned at the beginning of
\S\ref{sec:sims}, \citet{rah16} showed an under-estimate of the
frequency of $N_{\OVI}>10^{14.5} \cms$ absorbers at $z\approx 0-0.5$
in their Figure 2.  However, this result used the lower resolution
Ref-L100N1504 EAGLE volume, and we use the Recal-L025N0752 EAGLE
resolution and prescription for our {\it M5.3} zooms.  Opp16
demonstrated that $\OVI$ results are statistically identical between
the {\it M5.3} NEQ zooms (i.e. the No-AGN-Rad model) and the
Recal-L025N0752 volume.  \citet{rah16} tested resolution convergence,
finding $3\times$ more $N_{\OVI}=10^{14.5} \cms$ absorbers in the
Recal-L025N0752 volume (their Figure B2), which better agrees with the
observations compiled in their Figure 2.  Opp16 showed that the {\it
  M5.3} zooms produced $\OVI$ columns $0.26$ dex higher in the $L^*$
CGM than the lower resolution EAGLE volumes, and argued that this also
supported better agreement with high-column $\OVI$ IGM statistics.
PZFs around $L^*$ galaxies should raise the frequency of high-column
$\OVI$ absorbers even more, but their frequency in an IGM $\OVI$
survey like \citet{dan16} is low.  In summary, PZFs are unlikely to
contribute much to the observed $\OVI$ IGM statistics, and the level
of agreement between the \citet{rah16} Recal-L025N0752 volume results
and the IGM $\OVI$ observations remains approximately the same.

{\bf Spectral hardness of AGN spectrum:} We assume the \citet{saz04}
spectrum, which uses an extreme-UV (EUV) slope of $\nu^{-1.7}$ below
$1216$ \AA, which is slightly harder than the assumed \citet{haa01}
EGB.  This spectrum was derived for the average Type I unobscured
quasar from a variety of observations available at that time.  More
recent work by \citet{ste14} found an EUV slope of $\nu^{-1.41}$,
which is even harder and would create an even greater PZF effect for
high ions like $\OVI$ as discussed in OS13.  Observing the EUV slope
of Seyfert-like galaxies is critical for understanding the magnitude
of the PZF effect around star-forming galaxies, but we note that our
assumption of the \citet{saz04} spectrum could be a conservative
assumption for the PZF effect.

\section{Summary} \label{sec:summary}

AGN proximity zone fossils (PZFs) ionizing their CGM were introduced
by \citet{opp13b}, initially as a test of non-equilibrium ionization
and cooling in the presence of rapidly changing radiation fields.
\citet{seg17} ran the first cosmological hydro simulations of PZFs
using individual EAGLE haloes, finding high ions are significantly
enhanced for a variety of parameters.  Here, we argue that PZFs
significantly enhance $\OVI$ observed by COS-Halos around normal
star-forming galaxies \citep{tum11} using the first dynamic
simulations including PZFs.  AGN ionize the metal-enriched CGM around
typical star-forming disk galaxies and then turn off leaving CGM metal
ions far out of equilibrium for time scales longer than the time
between subsequent AGN episodes.  The central requirement for a
significant PZF effect is therefore that the star-forming galaxy
hosted an active AGN within a timescale comparable to the
recombination time of a high metal ion, which for circumgalactic
$\OVI$ is of the order of $10^7$ years.

We vary AGN lifetimes, luminosities, duty cycle fractions, and
isotropy in an EAGLE zoom simulation of a typical star-forming galaxy
using the non-equilibrium module introduced in \citet{opp16a} with the
addition of a spatially and time variable AGN field\footnote{Please
  visit http://noneq.strw.leidenuniv.nl/PZF/ for visualizations of the
  PZF effect.}.  Our focus is to investigate whether an AGN can
reproduce the strong $\OVI$ around star-forming galaxies observed by
COS-Halos while having an evolutionarily sustainable SMBH growth rate
and a small duty cycle fraction so that most galaxies appear as
inactive AGN.  We find that models with AGN bolometric luminosities
$\ga 10^{43.6} \ergs$, duty cycle fractions $\la 10\%$, and a
time-averaged power $\langle L_{\rm bol}\rangle \ga 10^{42.6} \ergs$
satisfy these requirements.  Critically, $\HI$ does not show the PZF
effect, because high ionization fraction enables it to quickly return
to equilibrium with the extra-galactic background, resulting in strong
$\HI$ column densities as observed around COS-Halos \citep{tho12}.
Our favoured models have specific SMBH accretion rates $\approx
2\times$ higher than the specific SFR of a typical star-forming
galaxy, which is consistent with cosmologically expected growth rates
of typical low-mass SMBHs occupying disk galaxies with $M_*\sim
10^{10}-10^{10.5} \msolar$.  Interestingly, recent surveys of
low-redshift, star-forming galaxies of this mass \citep{air17} find
X-ray-selected AGN with luminosities, duty cycle fractions, and
time-averaged powers fitting our requirements for significant PZF
effects.

AGN lifetimes are more difficult to constrain observationally, but
recent evidence of flickering AGN with timescales of $10^5$ yr or less
around star-forming galaxies in the evolved Universe
(\S\ref{sec:design} and \S\ref{sec:discuss}) support our argument for
non-equilibrium PZFs in the low-redshift CGM.  Models with $\tAGN\la
10^6$ yr can enhance $\OVI$ columns by the factors of $2-3$ necessary
to match COS-Halos.  Timescales $\la 10^5$ yr result in the CGM
ionization becoming temporally decorrelated from AGN activity,
resulting in a quasi-steady-state PZF with the ionization level
described by the time-averaged AGN+EGB field (Seg17).  Additionally,
we show in \S\ref{sec:lowions} that such models can create narrow
$\OVI$ absorbers that are well-aligned with low and intermediate ions
as observed by \citet{wer16}.  The dynamics of the CGM gas appear
unaltered between the model with no AGN radiation (No-AGN-Rad) and the
AGN PZF runs, which agrees with Seg17 that AGN photo-heating does not
significantly alter cooling in the CGM.

We argue that PZFs are a necessary consideration when deriving and
interpreting physical conditions of the highly ionized CGM probed by
the Cosmic Origins Spectrograph.  Our model presents a plausible
solution for the under-estimates of $\OVI$ column densities reported
by several teams analyzing the range of cosmological hydrodynamic
simulations of galaxy formation and evolution.  If normal disk
galaxies in the evolved Universe grow their SMBH at the cosmologically
expected rate, the PZF effect is likely to affect observations of the
metal-enriched CGM.

\section*{acknowledgments}

We are grateful for the anonymous referee who provided an insightful
review that extended our exploration in this manuscript.  The authors
would also like to thank Daniel Angles-Alcazar, Trystyn Berg,
Sebastiano Cantalupo, Julie Comerford, Sara Ellison, Mary Putman,
Jonathan Trump, Jess Werk, and Nadia Zakamska for useful discussions
contributing to this manuscript.  Support for Oppenheimer was provided
through the NASA ATP grant NNX16AB31G.  This work was supported by the
European Research Council under the European Union’s Seventh Framework
Programme (FP7/2007- 2013)/ERC Grant agreement
278594-GasAroundGalaxies and by the Netherlands Organisation for
Scientific Research (NWO) through VICI grant 639.043.409.  This work
used the DiRAC Data Centric system at Durham University, operated by
the Institute for Computational Cosmology on behalf of the STFC DiRAC
HPC Facility (www.dirac.ac.uk). This equipment was funded by BIS
National E-infrastructure capital grant ST/K00042X/1, STFC capital
grants ST/H008519/1 and ST/K00087X/1, STFC DiRAC Operations grant
ST/K003267/1 and Durham University. DiRAC is part of the National
E-Infrastructure.  RAC is a Royal Society University Research Fellow.
AJR is supported by the Lindheimer Fellowship at Northwestern
University.

\appendix

\section{Resolution test} \label{sec:res}

\begin{figure}
\includegraphics[width=0.49\textwidth]{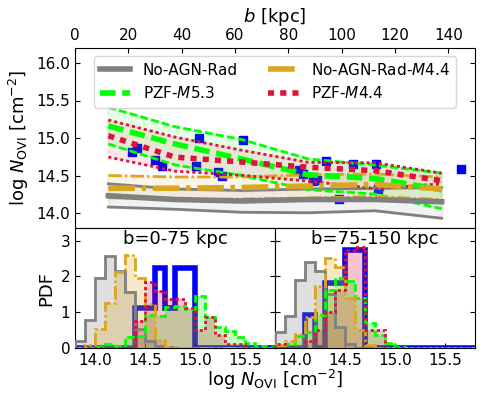}
\caption[]{SMOHALOS realizations are plotted as in
  Fig. \ref{fig:b_OVI_tAGN} using $\tAGN=10^5$ yr models ({\it
    L441d10t5}), for the fiducial {\it M5.3} resolution and the
  $8\times$ higher {\it M4.4} mass resolution models.}
\label{fig:b_OVI_resolution}
\end{figure}

We examine the {\it M4.4} runs with $8 \times$ higher mass resolution.
The No-AGN-Rad-{\it M4.4} run in Figure \ref{fig:b_OVI_resolution} has
0.17 dex higher columns than the {\it M5.3} resolution run, but the
$\OVI$ columns remain nearly a factor of $2$ too low.  Opp16 (\S6.1)
argued that the shortfall in simulated $\OVI$ might be rectified by
reducing the feedback efficiency to form more stars and hence
synthesize more oxygen, but this clearly does not work.  Our {\it
  M4.4} run has an $0.30$ dex higher stellar mass than the EAGLE-Recal
prescription {\it M4.4} in Opp16, but the $\OVI$ is nearly the same
level as this previous run.  

The {\it L441d10t6-M4.4} and {\it L441d10t5-M4.4} runs listed in Table
\ref{tab:PZFmodels} show milder ($\approx 2/3$rd as strong) increases
in $\OVI$ due to the PZF effect, but {\it L441d10t5-M4.4} provides an
excellent fit to the COS-Halos data, and closely follows the {\it
  L441d10t5} model in Fig. \ref{fig:b_OVI_resolution} (cf. maroon
dotted and green dashed lines).  The smaller PZF effect at higher
resolution owes to the weaker feedback that our {\it M4.4} model uses,
leaving CGM metals at a slightly higher density, and reducing the
strength of the PZF effect due to shorter recombination times.


\section{AGN-on models} \label{sec:AGNon}

\begin{figure}
\includegraphics[width=0.49\textwidth]{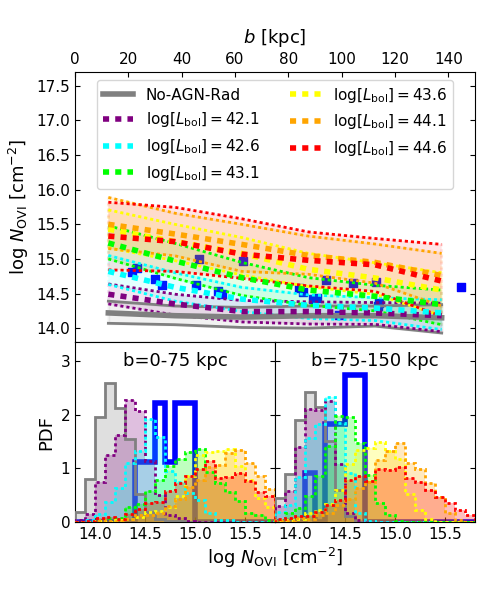}
\caption[]{SMOHALOS realizations are plotted as in
  Fig. \ref{fig:b_OVI_tAGN} using constant AGN-on models.  Increasing
  AGN radiation increases $\OVI$ column density.  }
\label{fig:b_OVI_AGNon}
\end{figure}

The effect of constant AGN irradiation is demonstrated in Figure
\ref{fig:b_OVI_AGNon} using the fiducial {\it M5.3} resolution.  Six
AGN-on models are shown increasing from $L_{\rm bol}=10^{42.1}$ to
$10^{44.6} \ergs$ in steps of $0.5$ dex.  All models show an increase
over the No-AGN-Rad model.  These models are more applicable to cases
when the AGN is observed to be on as in the COS-AGN observing program
(Berg et al., in prep).  The $L_{\rm bol}=10^{43.1} \ergs$ model has
the same time-averaged $\sdotMBH=2.5\times 10^{-10} {\rm yr}^{-1}$ as
the {\it L441d10t5} model, and almost exactly replicates its $\OVI$
statistics, since that PZF model is in the limit of AGN interval times
$\ll t_{\rm rec}$ where the column densities approach the steady state
of constant ionization by the AGN+EGB field.  

The increase of $\OVI$ saturates above
$L_{\rm bol} \approx 10^{43.6} \ergs$, because the AGN begins to
ionize more oxygen to higher levels ($\OVII$ through $\OIX$).  Even
higher AGN luminosities than those explored here (i.e. in the realm of
proper quasars) will result in less $\OVI$ than these peak models
(OS13).

\end{document}